\documentclass[a4paper,fleqn,usenatbib]{mnras}
\usepackage{latexsym}
\usepackage{amsmath}
\usepackage{amssymb}
\usepackage{fnpos}
\usepackage{hyperref}
\usepackage{tabularx}
\usepackage{xspace}
\usepackage{enumerate}
\usepackage{scalerel}
\usepackage{graphicx}
\usepackage{url}
\usepackage{caption}
\usepackage{gensymb}
\usepackage{longtable}
\usepackage{lscape}
\usepackage[normalem]{ulem} % To strike out text
\usepackage[dvipsnames]{xcolor} % To get a wider range of font colours
\usepackage{wasysym} %Better solar symbol, etc
\definecolor{darkgreen}{rgb}{0.0,0.55,0.0}
\definecolor{darkblue}{rgb}{0.0,0.0,0.5}

\newcommand{\eal}[2]{\ifmmode{\mathrm{#1\,#2}}\else{#1\textsc{$\,$\lowercase{#2}}}\fi\xspace}
\newcommand{\feal}[2]{\ifmmode{\mathrm{#1\,#2}}\else{[#1\textsc{$\,$\lowercase{#2}}]}\fi\xspace}
\newcommand{\hfeal}[2]{\ifmmode{\mathrm{#1\,#2}}\else{#1\textsc{$\,$\lowercase{#2}}]}\fi\xspace}

\newcommand{\orcid}[1]{$^{\rm \href{https://orcid.org/#1}{\includegraphics[height=0.6em]{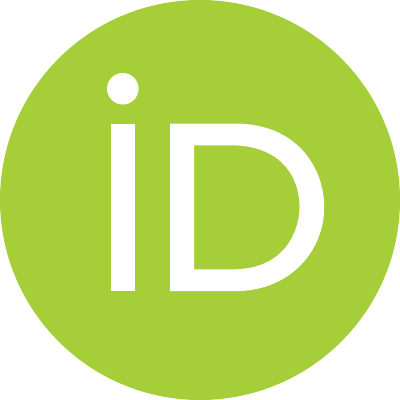}}}$}

\title[The flaring nova V612 Scuti]{From Light to Sound: Spectroscopic Evolution \& Sonification of the flaring Nova V612~Scuti}
\author[Pragati et al.]{\parbox{\textwidth} {P.~Acharya$^{1}$, E. Aydi\orcid{0000-0001-8525-3442}$^{1}$\thanks{E.~Aydi; E-mail: eaydi@ttu.edu}, K.~V.~Sokolovsky\orcid{0000-0001-5991-6863}$^{1}$, L.~Chomiuk\orcid{0000-0002-8400-3705}$^{2}$, P.~Craig$^{2}$, L.~Izzo\orcid{0000-0001-9695-8472}$^{3}$, J.~D.~Linford\orcid{0000-0002-3873-5497}$^{4}$, B.~D.~Metzger\orcid{0000-0002-4670-7509}$^{5, 6}$, S. Mohamed$^{7-11}$, I. Molina$^{2}$, K.~Mukai$^{12, 13}$, K.~J.~Shen$^{14}$, J.~L.~Sokoloski$^{5}$, P.~Luckas$^{15}$, T.~Bohlsen\orcid{0000-0003-0371-2639}$^{16}$, O.~Garde\orcid{0000-0002-7850-8360}$^{17}$, J.~G.~Fl\'{o}$^{16,18,19}$, P.~Bellavista\orcid{0000-0003-0992-7948}$^{20}$, S.~Umberto$^{21}$, and F.~Sims$^{22}$}
\vspace{0.4cm}\\
\parbox{\textwidth}{
$^{1}$Department of Physics and Astronomy, Texas Tech University, Lubbock, TX 79409, USA\\
$^{2}$Center for Data Intensive and Time Domain Astronomy, Department of Physics and Astronomy, Michigan State University, East Lansing, MI 48824, USA\\
$^{3}$INAF-Osservatorio Astronomico di Capodimonte, Salita Moiariello 16, 80131, Napoli, Italy\\
$^{4}$National Radio Astronomy Observatory, P.O.\ Box O, Socorro, NM 87801, USA\\
$^{5}$Department of Physics and Columbia Astrophysics Laboratory, Columbia University, New York, NY 10027, USA\\
$^{6}$Center for Computational Astrophysics, Flatiron Institute, 162 5th Ave, New York, NY 10010, USA\\
$^{7}${Department of Astronomy, University of Virginia, 530 McCormick Road, Charlottesville, VA 22904, USA}\\
$^{8}${Virginia Institute for Theoretical Astronomy, University of Virginia, Charlottesville, VA 22904, USA}\\
$^{9}${South African Astronomical Observatory, P.O Box 9, Observatory, 7935, Cape Town, South Africa}\\
$^{10}${Department of Astronomy, University of Cape Town, Private Bag X3, Rondebosch, 7701, Cape Town, South Africa}\\
$^{11}${NITheCS National Institute for Theoretical and Computational Sciences, South Africa}\\
$^{12}$CRESST and X-ray Astrophysics Laboratory, NASA/GSFC, Greenbelt, MD 20771, USA\\
$^{13}$Department of Physics, University of Maryland, Baltimore County, 1000 Hilltop Circle, Baltimore, MD 21250, USA\\
$^{14}$Department of Astronomy and Theoretical Astrophysics Center, University of California, Berkeley, CA 94720, US\\
$^{15}$International Centre for Radio Astronomy Research, The University of Western Australia, 35 Stirling Hwy Crawley, Western Australia 6009, Australia\\
$^{16}$Astronomical Ring for Access to Spectroscopy\\
$^{17}$Observatoire de la Tourbi\`{e}re 45, Chemin du Lac 38690 CHABONS, France\\
$^{18}$Piera Remote Observatory, J.Balmes,2   08784 PIERA, Catalonia\\
$^{19}$SMM Remote Observatory, Av, Catalunya,38 25354, Santa Maria de Montmagastrell,  Catalonia\\
$^{20}$Dept. Computer Science and Engineering (DISI), University of Bologna, Viale del Risorgimento, 2 - 40136 Bologna - Italy\\
$^{21}$Via San Sisto, 20 67100 L'Aquila - Private observatory, Italy\\
$^{22}$Desert Celestial Observatory, 5900 E Bell St. Apache Junction, Arizona 85119, USA\\
}}

\begin{document}
\label{firstpage}
\pagerange{\pageref{firstpage}--\pageref{lastpage}}
\maketitle

\begin{abstract}
We present photometric and spectroscopic observations of the 2017 Galactic nova V612~Sct, whose optical evolution was marked by multiple unusually large maxima. The eruption included two prominent flares lasting around a month each and reaching amplitudes of about 2.5 mag, followed by a series of smaller flares. Extensive spectroscopic monitoring reveals a striking pattern: with each flare, new absorption systems emerge at progressively higher velocities. This behavior, also seen in other flaring novae, provides evidence for repeated episodes of mass ejection or outflow at increasing velocities. V612~Sct also alternated between Fe~II and He/N spectral phases during different stages of the eruption, establishing a clear connection between the photometric flares and major spectral transitions. We present two-dimensional dynamic spectra that directly trace the appearance of new absorption features contemporaneous with the light-curve flares. We also introduce a sonification of the spectroscopic sequence, offering an alternative representation of the temporal evolution of the eruption. These results support a picture in which repeated ejection episodes and shock formation play a central role in powering the multiple maxima observed in flaring novae.
\end{abstract}


\section{Introduction}

Classical novae are explosive transient events that occur in close binary systems, where a white dwarf (WD) accretes hydrogen-rich material from a companion star (see \citealt{Bode_etal_2008,Della_Valle_Izzo_2020,Chomiuk_etal_2021} for a review). Once a critical mass has been accreted, a thermonuclear runaway is triggered on the surface of the white dwarf. This leads to the ejection of material at high velocities (hundreds to thousands of km\,s$^{-1}$) and a dramatic brightening, typically by 8--15 magnitudes, producing a characteristic light curve that rises rapidly to maximum and then declines smoothly as the photosphere recedes and the emission shifts blueward into the ultraviolet and eventually X-ray regimes \citep{Gallagher_Starrfield_1976,Starrfield_1989_bode,Strope_etal_2010,Aydi_etal_2020b}. In the traditional picture, the luminosity is attributed to reprocessed emission from residual nuclear burning on the white dwarf's surface following the eruption \citep{Starrfield_1989, Wolf_etal_2013}. However, more recent studies have established that strong shocks resulting from the interaction of multiple phases of mass loss can drive a substantial fraction of the visible emission in novae. This interpretation is supported by the detection of $\gamma$-ray emission in several novae with \textit{Fermi}-LAT \citep{Ackermann_etal_2014,Cheung_etal_2016,Franckowiak_etal_2018}, which in some cases is directly correlated with the optical emission, linking shock interactions to the optical output \citep{Li_etal_2017_nature,Aydi_etal_2020a,Chomiuk_etal_2021}.

While the optical light curves of some novae show a smooth decline, a subset known as flaring novae exhibit erratic peaks (``flares'' or ``jitters'') superposed on the broader light curve (see e.g., \citealt{Strope_etal_2010}). These flares/jitters could show a brightness increase for a few days up to a few weeks, while their origin remain debated. Explanations include instabilities/pulsations in the envelope of the WD \citep{Schenker_1999,Pejcha_2009}, renewed mass ejection episodes \citep{Pejcha_2009,Aydi_etal_2019_I}, and instabilities in a large accretion disk that survived the eruption \citep{Goranskij_etal_2007}. With the new insight brought by the now-routine detection of GeV $\gamma$-rays from Galactic novae, mounting evidence points to a central role for internal shock interactions (faster ejecta overtaking earlier, slower-moving material) in powering substantial and enhanced visible emission in novae \citep{Li_etal_2017_nature,Aydi_etal_2020a}. Notably, the 2018 nova V906 Carinae revealed compelling direct evidence for shock-powered flares \citep{Aydi_etal_2020a}. Simultaneous optical and $\gamma$-ray observations demonstrated tightly correlated light-curve peaks, indicating that the flares originate in shocks rather than white dwarf surface luminosity. %Hard X-ray observations further revealed that much of the shock energy does not appear in X-rays, implying absorption and reprocessing of shock power into optical wavelengths \citep{Aydi_etal_2020a,Sokolovsky_etal_2020}. 
These findings support a broader picture in which internal shocks substantially contribute to the bolometric emission in novae \citep{Chomiuk_etal_2021}. A closely analogous event is the very slow nova ASASSN-17pf (LMCN 2017-11a), studied by \citet{Aydi_etal_2019_I}, where pronounced flares, the onset of dust formation, and spectroscopic evidence of shock interactions were observed---including multiple absorption features at increasing velocities.

Here, we study the 2017 nova V612~Sct (ASASS-17hx), which was discovered by the All-Sky Automated Survey for Supernovae (ASAS-SN), on UT 2017-06-23.47, and first detected in eruption on 2017-06-19.41 \citep{ATel_10523,ATel_10527}. We use 2017-06-19.41 (HJD 2457923.91) as $t_0$ for the remainder of the paper. The eruption of V612~Sct was notable for two long-lasting ($\sim$ weeks-scale) flares with amplitudes of $\approx$ 2.5 magnitudes, followed by a series of smaller flares before disappearing behind the Sun for $\sim$120 days, after which it was recovered on the decline. Due to its high-brightness and slow evolution, the nova was extensively studied \citep{Rosenthal_etal_2018,Chochol_etal_2019,Mason_etal_2020,Rudy_etal_2024} and observed spectroscopically by citizen scientists during the flaring period, allowing us to study the spectroscopic evolution of the event in exquisite detail. In this paper, we present our comprehensive analysis of the photometric and spectroscopic observations of V612~Sct. Section \ref{sec:observations} describes the spectro-photometric data. In Section \ref{sec:results}, we detail the light-curve morphology, spectral-line profile evolution, 2D dynamic spectra, and our sonification approach for the spectroscopic time series. Finally, in Section \ref{sec:discussion}, we interpret these findings in the context of multiple phases of mass ejection, shock interaction, and the origins of multiple maxima in flaring novae, followed by our conclusions.

\section{Observations}
\label{sec:observations}

We make use of publicly available optical spectra from the Astronomical Ring for Access to Spectroscopy (ARAS; \citealt{Teyssier_2019}), which were obtained by a variety of observers from around the world. The data we adopt here are mostly high-resolution ($R \sim 10000$) and some low-resolution ($R \sim 1000$), covering the range 4000\,--7400\,\AA. The dataset spans around 390 days of observations covering the early rise, peak, flaring, and the decline phases of the eruption. All spectra were continuum-normalized using the Image Reduction and Analysis Facility (IRAF; \citealt{Tody_1986}) and exported as ASCII files for subsequent analysis. The log of the spectroscopic observations is presented in Tables~\ref{table:spec_log_V612_Sct_1} and~\ref{table:spec_log_V612_Sct_2}.

We also make use of publicly available photometry from the American Association of Variable Stars (AAVSO; \citealt{AAVSODATA}) International Database, consisting of CCD and CMOS photometry, mostly in the $V$-band, $B$-band, $R$-band, $I$-band,
and unfiltered band with $V$ zero-point ($CV$). These data were used to construct the optical light curve and to identify photometric phases corresponding to changes in the spectral line profiles. A sample of the photometric data is presented in Table~\ref{tab:photometry_sample}.

\section{Results}
\label{sec:results}

In this section we present the photometric and spectroscopic evolution of V612~Sct throughout its eruption. We begin with the optical light curve, which establishes the temporal framework for the nova's brightness evolution and decline rates. We then examine the color evolution, followed by a detailed analysis of the spectroscopic development, including changes in line profiles, the emergence of emission features, and the 
velocity evolution of the ejecta. Together, these results provide a coherent picture of the physical processes governing the nova throughout its outburst.

\begin{figure*}
\begin{center}
  \includegraphics[width=1.0\textwidth]{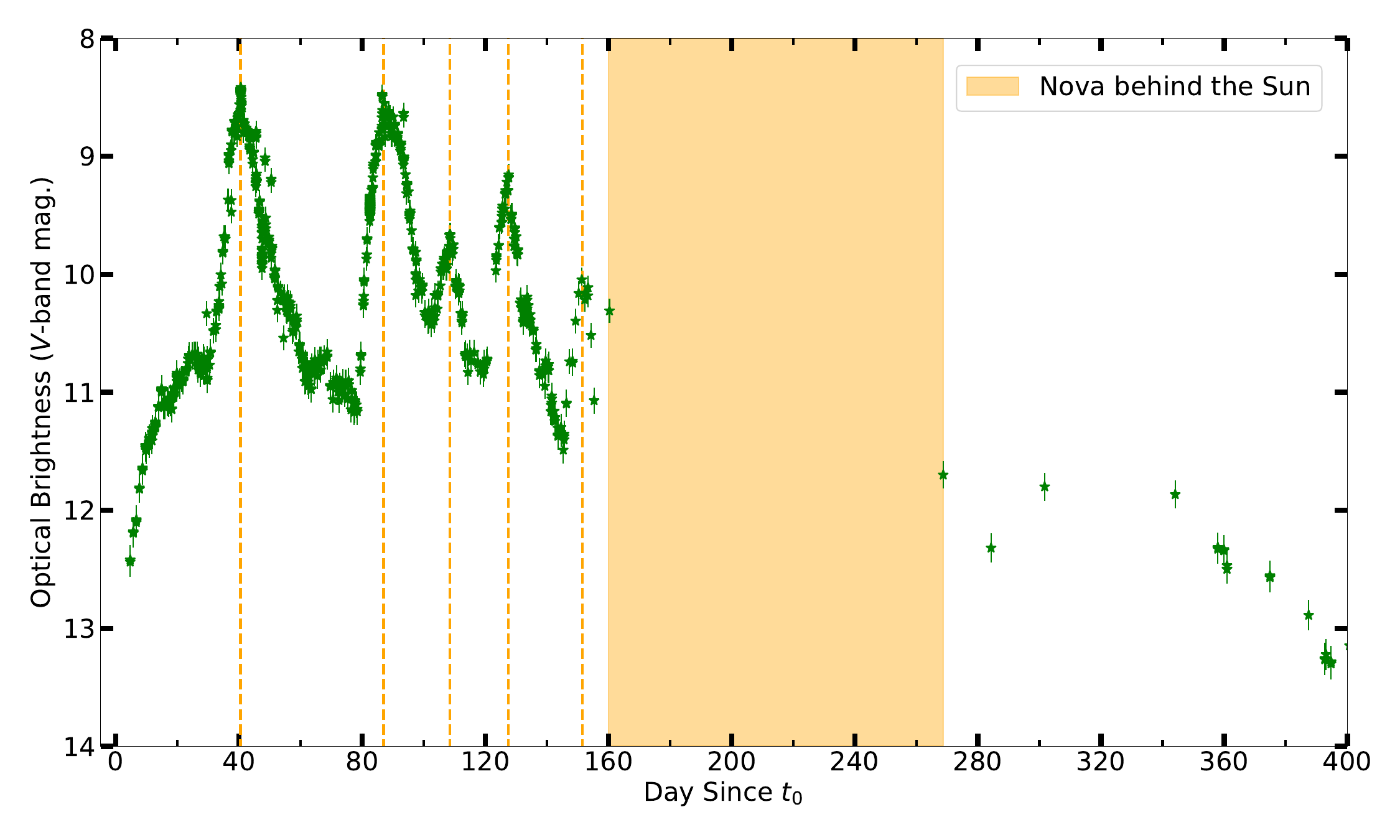}
\caption{The optical $V$-band light curve of nova V612~Sct. The orange dashed lines mark the peak of the flares (maxima), while the yellow shaded region denotes the period of solar conjunction. The error bars represent 1-$\sigma$ uncertainties.} 
\label{Fig:light_curve}
\end{center}
\end{figure*}

\subsection{Photometric Evolution}

Figure~\ref{Fig:light_curve} shows the AAVSO $V$-band light curve of V612~Sct, covering the first 400 days of the eruption. The nova reached its (first) maximum brightness of around 8.4~mag in the $V$-band on JD = 2457964.48 (around 40 days since $t_0$). We adopt this date as the time of maximum light ($t_{\mathrm{max}}$) in the $V$-band.

After the first maximum, the light curve exhibits at least four additional maxima (flares) with a range of amplitudes and timescales (orange dashed lines in Figure~\ref{Fig:light_curve}). These maxima are centered at approximately days 87, 108, 127, and 151 after $t_0$. We consider a flare/maximum a jump of flux by 2 times (or a decrease in the magnitude by 0.8\,mag). The properties of the flares are summarized in Table~\ref{Table:flares}. The two largest flares have durations of $\Delta t = 32$ and 24 days, respectively, and amplitudes of $\Delta V = 2.4-2.5$ mag, corresponding to an approximately tenfold increase in the optical luminosity of the nova. The flare duration is defined as the interval from the onset of the brightness increase above the baseline level of the light curve to the end of the decline, when the brightness returns to the baseline level, or until the next flare begins immediately. This represents a substantial brightness enhancement above the underlying baseline light curve. The three largest flares show faster rise rates ($d V_{\mathrm{rise}}$) than decline rates ($d V_{\mathrm{decline}}$), whereas the two smallest flares exhibit decline rates that are comparable to, or even faster than, their rise rates. The main flaring phase continues until approximately day~160, after which the nova passes behind the Sun for about 110 days. As a result, we cannot determine whether additional flaring activity occurred during that interval. The origin of these flares will be discussed in following sections in context of spectroscopic evolution, multiple phases of mass-loss, and shock formation. 

After solar conjunction, the nova was detected again in decline and continued to fade gradually. The last epoch at which the nova remained brighter than 10.4~mag occurred at approximately day 154, implying that $t_2$---the time required for the nova to fade by 2 magnitudes from peak brightness---is at least 114 days. If additional brightening episodes occurred during solar conjunction, then the true value of $t_2$ may be even longer. Regardless, V612~Sct can be confidently classified as a slow nova.

\begin{table}
\centering
\caption{Characteristics of the flares observed in the optical light curve of nova V612~Sct, including the time of peak brightness relative to $t_0$ ($t_{\mathrm{peak}} - t_0$), the flare duration ($\Delta t$ in days), the flare amplitude  ($\Delta V$ in magnitude), the rise rate ($dV_{\mathrm{rise}}$), and the decline rate ($dV_{\mathrm{decline}}$).}
\begin{tabular}{ccccc}
\hline
\shortstack{$t_{\mathrm{peak}}-t_0$\\(d)} &
\shortstack{$\Delta t$\\(d)} &
\shortstack{$\Delta V$\\(mag)} &
\shortstack{$d V_{\mathrm{rise}}$\\(mag d$^{-1}$)} &
\shortstack{$d V_{\mathrm{decline}}$\\(mag d$^{-1}$)}\\ 
\hline
40  & 32 & 2.4 & 0.23 & 0.11\\
87  & 24 & 2.5 & 0.28 & 0.18\\
108 & 12 & 0.8 & 0.12 & 0.15\\
127 & 12 & 1.7 & 0.23 & 0.16\\
151 & 10 & 1.5 & 0.25 & 0.33\\
\hline
\label{Table:flares}
\end{tabular}
\end{table}

\citet{Pejcha_2009} and \citet{Tanaka_etal_2011_159} studied several slow novae with multiple optical maxima and found that the intervals between successive flares often follow a systematic trend. In particular, \citet{Pejcha_2009} proposed that the spacing between maxima is approximately uniform on a logarithmic scale, such that the interval between two consecutive peaks obeys a power-law relation of the form
\begin{equation}
\label{equation_1}
\log\left[t_{i} - t_{i-1}\right] = a + b\,\log\left[t_{i} - t_{\mathrm{max}}\right],
\end{equation}
where $t_i - t_{i-1}$ is the time interval between two successive flares. For most novae in their sample, they found $b \approx 1$. In contrast, \citet{Aydi_etal_2019_I} found no evidence for such a relation in the optical light curve of the flaring Large Magellanic Cloud nova ASASSN-17pf, where the intervals between flares appeared to be random.

The peak spacing in V612~Sct shows an overall declining trend, followed by a slight increase between the last two observed flares. The intervals between successive maxima are $(t_2-t_1)=47$~days, $(t_3-t_2)=21$~days, $(t_4-t_3)=19$~days, and $(t_5-t_4)=24$~days. Thus, the flare spacing in V612~Sct does not appear to follow the trend reported by \citet{Pejcha_2009}. %In that work, the intervals between successive rebrightenings in most novae were found to increase with time and to be approximately equally spaced in logarithmic time, corresponding to a positive power-law index typically close to unity.
In contrast, fitting the flare timings of V612~Sct with equation~\ref{equation_1} yields $a \approx 2.96$ and $b \approx -0.83$. The negative value of $b$ indicates that the flare intervals in V612~Sct do not systematically increase with time, and show an overall decrease with time.

The flare durations, however, decrease with time: the first flare lasts slightly more than a month, whereas the last flare persists for only about 10 days before the nova passes behind the Sun. The flare amplitudes do not show a clear systematic trend. Although the first two flares have the largest amplitudes, the subsequent flares exhibit more irregular variations in amplitude.

After solar conjunction, the photometric monitoring decreased significantly, however, data from the All-Sky Survey for Supernovae (ASAS-SN; \citealt{Shappee_etal_2014,Kochanek_etal_2017}) show that the nova exhibits further variability with brightness increase by 0.5--1.0\,mag around 330 and 430 days after $t_0$ (see Figure~\ref{Fig:ASASSN_LC}).Similar late-time variability has been observed in several other novae, including the recurrent rebrightenings of nova V4745~Sgr, the pronounced secondary brightening of nova V2362~Cyg, and the sub-magnitude irregular variability observed during the prolonged post-eruption plateau of nova V5856~Sgr \citep{Csak_etal_2005,Kimeswenger_etal_2008,Munari_etal_2022}. Proposed explanations for such behavior range from instabilities in residual hydrogen-shell burning and temporary re-expansion of the nova photosphere to renewed mass loss or secondary ejection episodes \citep{Kimeswenger_etal_2008,Pejcha_2009,Strope_etal_2010,Tanaka_etal_2011}. In the absence of contemporaneous spectroscopy at these late epochs, however, the physical origin of the additional brightenings in V612~Sct remains uncertain.

\begin{figure*}
\begin{center}
  \includegraphics[width=1.0\textwidth]{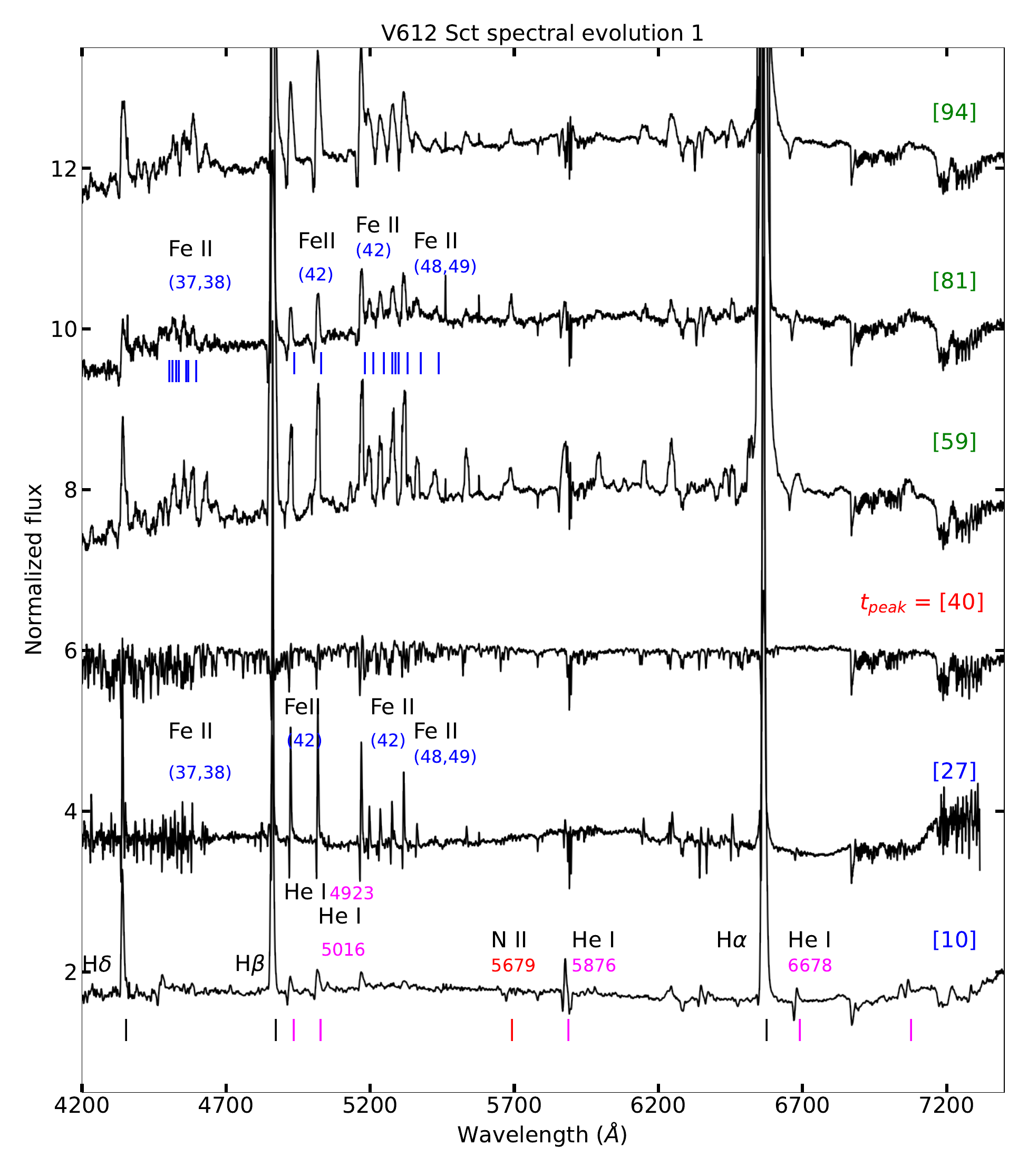}
\caption{The optical spectral evolution of V612~Sct during different stages of its outburst. The numbers in brackets are days since $t_0$ (these numbers are colored in blue for spectra taken before first peak and in green for spectra taken after first peak). Colored line identifications are also included to help distinguish the spectral features.} 
\label{Fig:main_spec_1}
\end{center}
\end{figure*}

\begin{figure*}
\begin{center}
  \includegraphics[width=1.0\textwidth]{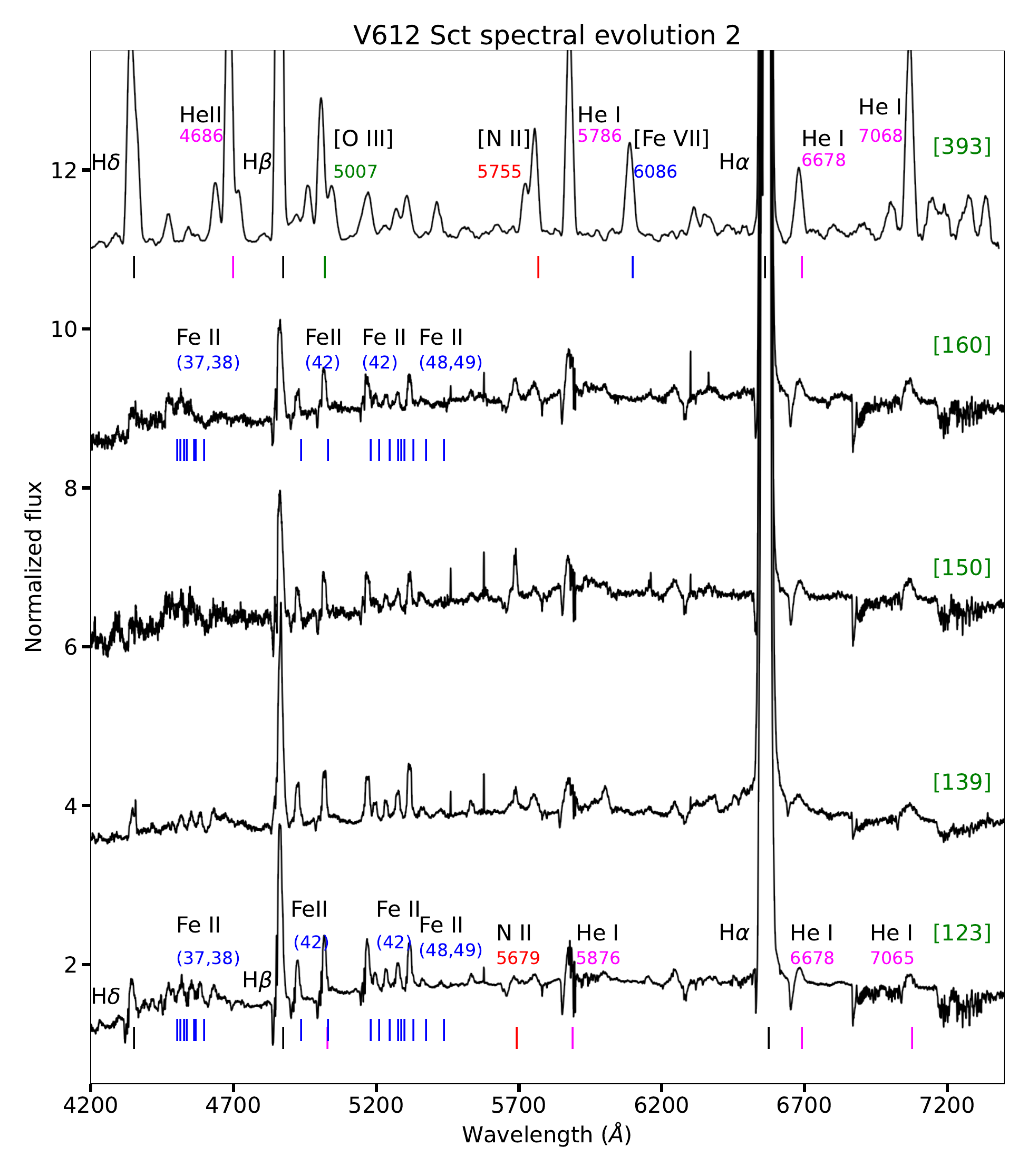}
\caption{Same as Figure~\ref{Fig:main_spec_1} -- continued.} 
\label{Fig:main_spec_2}
\end{center}
\end{figure*}

\subsection{Spectral evolution}

\begin{figure}
\begin{center}
  \includegraphics[width=1.0\columnwidth]{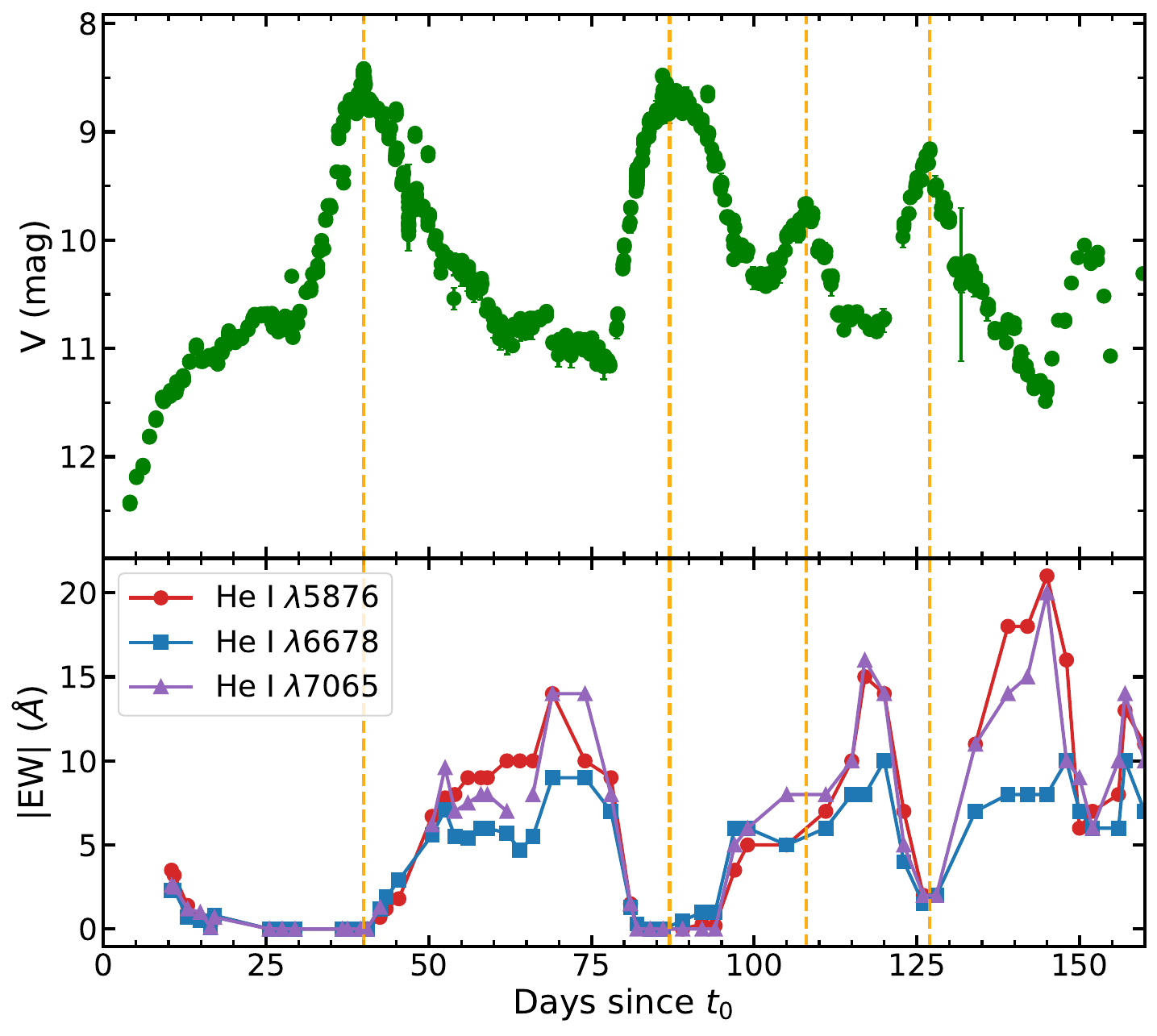}
\caption{The $V$-band light curve (\textit{top}) and the absolute equivalent widths, $|\mathrm{EW}|$, of the He I $\lambda5876$, $\lambda6678$, and $\lambda7065$ emission lines (\textit{bottom}) for the nova V612~Sct. The vertical dashed lines mark the centers of the flares. Since emission lines have negative EWs by convention, their absolute values are shown here.} 
\label{Fig:He_LC}
\end{center}
\end{figure}

Figures~\ref{Fig:main_spec_1} and~\ref{Fig:main_spec_2} show the overall spectral evolution of nova V612~Sct, while the detailed evolution is presented in Figures~\ref{Fig:spec_1} through~\ref{Fig:spec_13}. The first optical spectrum, obtained on day 10, approximately 30 days before maximum light, shows strong emission lines of Balmer, He~I, along with weaker N~II lines. At this stage, the spectrum is characteristic of the early He/N phase described by \citet{Williams_2012,Aydi_etal_2024}. A few days later, the Fe~II lines from multiplets (42), (48), and (49) strengthened, while the He and N lines weakened, indicating that the nova had transitioned into the Fe~II phase. \citet{Aydi_etal_2024} showed that most novae evolve through at least three distinct spectroscopic phases: He/N (1) \(\rightarrow\) Fe~II \(\rightarrow\) He/N (2), before entering the nebular phase. This evolution is thought to be associated with changes in the optical depth and ionization state of the ejecta \citep{Williams_2012,Shore_2014}. Around day 66 (Figure~\ref{Fig:spec_5}), prior to the second flare, the He/N lines strengthened again, indicating that the nova was entering the second He/N phase. Remarkably, during the second flare (between days 80 and 90), the Fe~II lines became stronger once more (Figure~\ref{Fig:spec_6}), while the He/N lines weakened again. This pattern repeats with each major flare; for example before the major flare on day 127 (flare 4), the He/N strengthen before weakening again during that flare while the Fe II lines strengthen (Figure~\ref{Fig:spec_9}). To highlight the evolution of the He emission-line strengths, we plot the equivalent widths (EWs) of He I 5876, 6678, and 7065,\AA\ alongside the optical light curve in Figure~\ref{Fig:He_LC}. The He I lines weaken dramatically during the flares, in some cases nearly disappearing, before strengthening again during the intervals between successive flares. Although a decrease in EW is expected during a flare because of the enhanced continuum emission, the near-complete disappearance of the He I lines points to changes in the ionization state of the ejecta. \citet{Aydi_etal_2024} and \citet{Aydi_etal_2026} showed that flaring novae, that is, novae with multiple maxima, can alternate repeatedly between the He/N and Fe~II phases in step with the flares in their optical light curves. This behavior implies that significant changes in the ionization state of the ejecta occur during the flaring activity. Detailed discussion of the ionization change in the spectra of nova V612~Sct are presented in \citet{Mason_etal_2020}.

The spectra of V612~Sct were dominated by emission lines or P~Cygni profiles of Balmer,  Fe~II and/or He/N lines up to solar conjunction, around day 160 (see Figures~\ref{Fig:spec_1} through~\ref{Fig:spec_13}). Additional spectroscopic observations were not obtained until approximately 360 days after eruption, a few months after the end of solar conjunction. By that stage, the spectra show forbidden emission lines of O~III, O~I, and N~II, along with He~II and He~I lines, indicating that the nova is in the nebular phase (Figure~\ref{Fig:spec_13}).  

\subsection{Line profile evolution}

\begin{figure*}
\begin{center}\includegraphics[width=\textwidth]{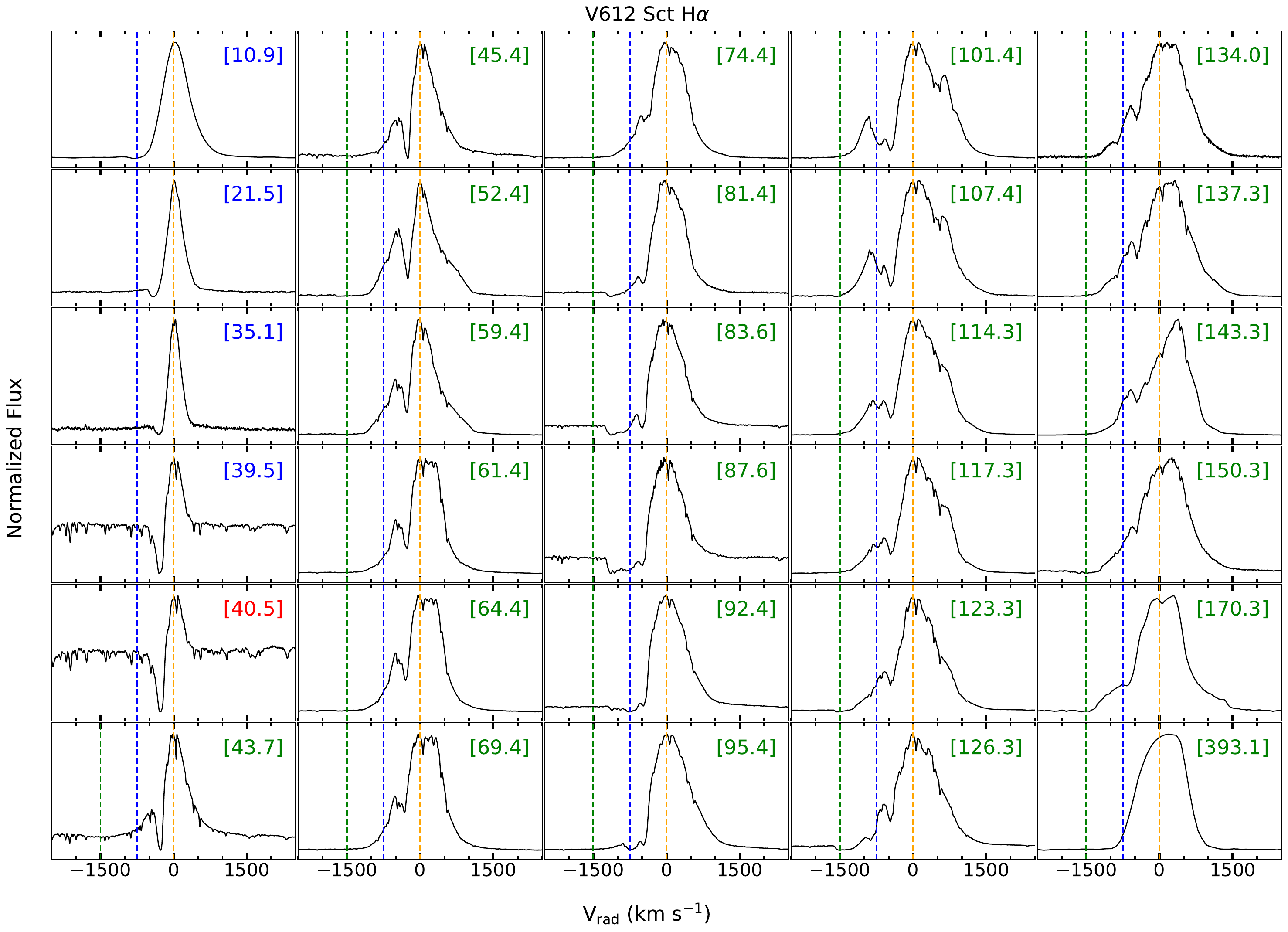}
\caption{The evolution of the  H$\alpha$ line profiles. The numbers between brackets are days since $t_0$. The days highlighted in blue are of spectra taken before peak brightness (the day highlighted in red represents peak brightness), while days highlighted in green are that of spectra taken after peak brightness. The orange, blue, and green dashed lines represent $v_{\mathrm{rad}}$ = 0\,km\,s$^{-1}$, $v_{\mathrm{rad}}$ = $-750$\,km\,s$^{-1}$, $v_{\mathrm{rad}}$ = $-1500$\,km\,s$^{-1}$, respectively. See text for more details.} 
\label{Fig:line_profile_main}
\end{center}
\end{figure*}

Figure~\ref{Fig:line_profile_main} presents a representative sample of the evolution of the H$\alpha$ line profile in velocity space over the course of the eruption. The full evolution of H$\alpha$ is shown in Figures~\ref{Fig:line_profile_1} through~\ref{Fig:line_profile_5}. The earliest spectra, obtained between days 10 and 14, exhibit weak P~Cygni absorption troughs at blueshifted velocities of approximately 750\,km\,s$^{-1}$. These troughs are not well resolved, owing to the spectral resolution, particularly because the earliest spectra were taken at relatively low resolution ($R \lesssim 1000$).

Around day 15, a P~Cygni absorption trough begins to emerge more clearly, with minima at blueshifted velocities of approximately 500\,km\,s$^{-1}$. However, these absorption features remain much weaker than the emission component; in other words, the emission lines are still relatively strong compared to the continuum. By days 35--39, a few days before peak brightness, the P~Cygni absorptions become very prominent and are characterized by trough minima at blueshifted velocities of about 300\,km\,s$^{-1}$. The changing relative strength of the emission and absorption components in the P~Cygni profiles is likely driven by the strengthening continuum as the nova rises toward peak brightness.

The apparent deceleration of the P~Cygni absorption profile prior to maximum light has been widely observed in nova spectra (e.g., \citealt{Aydi_etal_2020b}), and it is generally interpreted as a consequence of the decreasing optical depth of the ejecta as they expand, causing the photosphere to recede inward toward slower-moving layers, in a scenario similar to a Hubble flow \citep{Shore_2014}.

The weak absorption components (relative to the emission) in the P Cygni profiles during the early rise is also not uncommon, particularly for slowly rising novae (see e.g., \citealt{Tanaka_etal_2011,Aydi_etal_2026}). During the slow rise to peak, many novae show strong emission lines with shallow absorption features. \citet{Aydi_etal_2026} suggested that during this stage, the bulk of the nova ejecta is not yet fully ejected, and the nova is losing mass via low-density winds, which is responsible for the strong emission (relative to the continuum) and shallow absorptions. During this period, the bulk of the nova envelope is still engulfing the binary and expanding steadily, similar to a red giant star, until it is fully ejected, leading to the visible peak brightness. \citet{Aydi_etal_2026} relied on near-infrared interferometric imaging of the slow rising nova V1405~Cas, which shows little to no nova ejecta around the system during the rise to peak, and a resolved central source with the size of a red giant star. 

After the first optical maximum on day 40, a broader emission base becomes visible, with the pre-maximum P~Cygni profile superimposed on top of it. The emission wings extend to velocities of approximately 1500\,km\,s$^{-1}$. This behavior is common in novae \citep{McLaughlin_1942,Friedjung_1966_I,Friedjung_1966_III,Friedjung_2011}. \citet{Aydi_etal_2020b} showed that novae generally exhibit at least two velocity components during the rise to peak and the early decline: a slower component, traced by the low-velocity pre-maximum P~Cygni absorption, and a faster component, traced by the broad emission underlying it (see also, \citealt{McLaughlin_1942,Friedjung_1966_I,Friedjung_1966_III,Friedjung_2011}). \citet{Aydi_etal_2020b} interpreted these two components as signatures of two distinct outflows: an early slow flow followed by a faster wind. The fact that these different velocity components coincide within the same lines (i.e., the same species) implies the presence of multiple distinct origins, or outflows, characterized by different velocities.

\begin{figure}
\begin{center}
\includegraphics[width=\columnwidth]{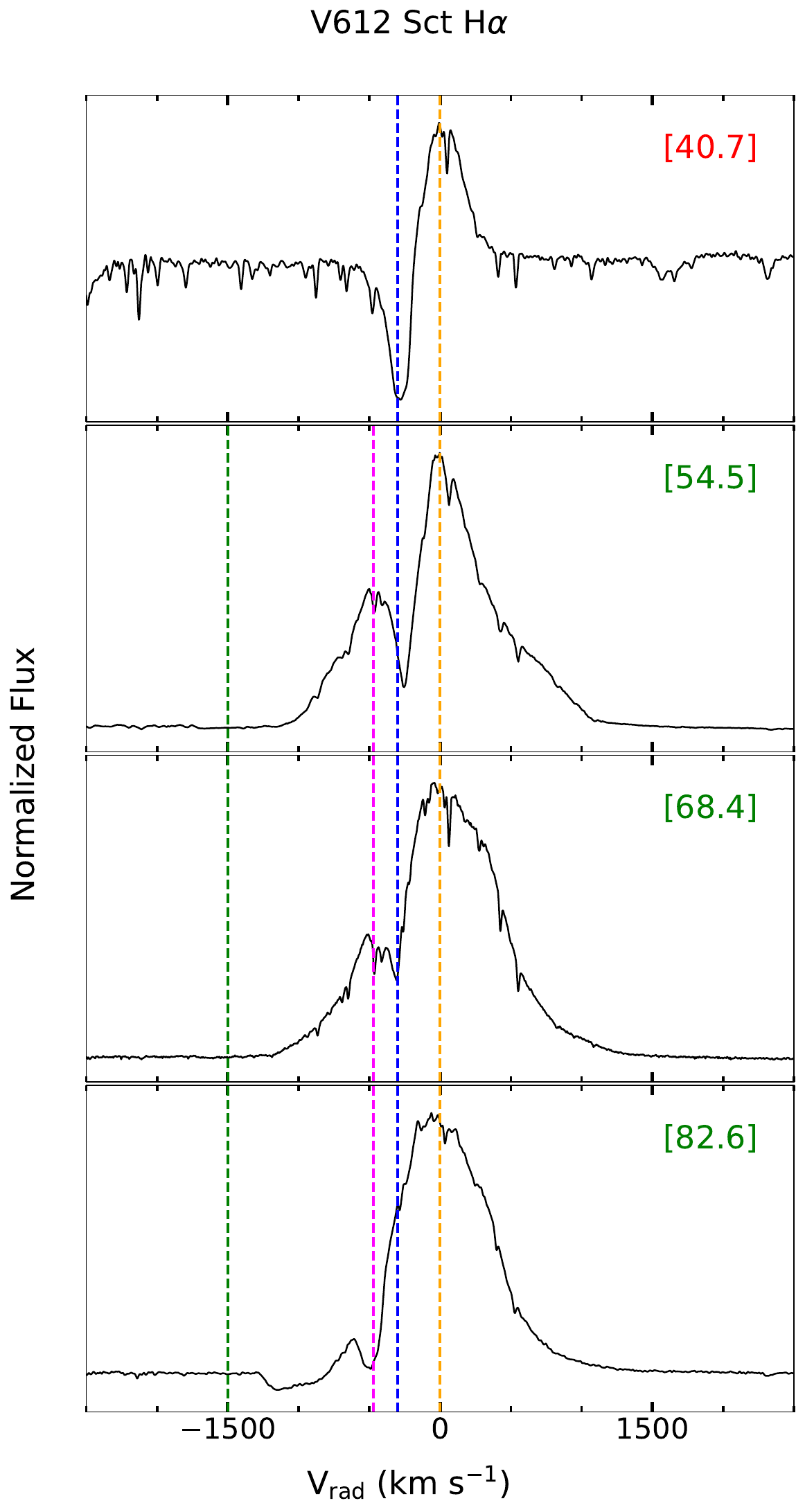}
\caption{H$\alpha$ line profiles showing the emergence of an intermediate velocity component between the slow and fast components. The orange, blue, magenta, and green dashed lines represent $v_{\mathrm{rad}}$ = 0\,km\,s$^{-1}$, $v_{\mathrm{rad}}$ = $-300$\,km\,s$^{-1}$, $v_{\mathrm{rad}}$ = $-500$\,km\,s$^{-1}$, $v_{\mathrm{rad}}$ = $-1500$\,km\,s$^{-1}$, respectively. See text for more details.}
\label{Fig:4_lines}
\end{center}
\end{figure}

In addition to the slower and faster components, an intermediate component, often referred to as the principal component \citep{McLaughlin_1942,McLaughlin_1944,Payne-Gaposchkin_1964,Friedjung_1987}, can also be identified after peak brightness when sufficiently dense spectroscopic coverage is available. Because V612~Sct was monitored extensively, we are able to detect the emergence of this intermediate component during the decline from the first maximum, at a velocity of approximately $-500$\,km\,s$^{-1}$. This component persists throughout the subsequent evolution of the nova, as is commonly observed for the principal component. \citet{Mason_etal_2020} also identified this velocity component in V612~Sct and noted its persistence from the early decline onward. They ruled out an origin in circumbinary material and instead suggested that it is associated with a shell of gas within the nova ejecta. \citet{Friedjung_1987} and \citet{Aydi_etal_2020b} proposed that the principal component arises from the interaction between the slower pre-maximum flow and the later, faster flow (see figure 12 in \citealt{Aydi_etal_2020b}). This interaction produces shocks, which give rise to high-energy emission, and forms an expanding shell that sweeps through the slower ejecta. As this shell propagates through the slow flow, the slower pre-maximum absorption component gradually disappears and is replaced by the intermediate component (See figures 8 and 9 in \citealt{Aydi_etal_2020b}). This behavior is also observed in the H$\alpha$ spectral-evolution sequence of V612~Sct (Figures~\ref{Fig:line_profile_1}--\ref{Fig:line_profile_5}). To illustrate this more clearly, we highlight a few selected epochs in Figure~\ref{Fig:4_lines}, emphasizing the slower, faster, and intermediate components. The spectrum obtained around day 40, near peak brightness, shows a strong absorption feature at $-300$\,km\,s$^{-1}$ (blue dashed line). About two weeks later, a new absorption feature appears at $-500$\,km\,s$^{-1}$ (magenta dashed line); this feature strengthens with time and eventually overtakes the component at $-300$\,km\,s$^{-1}$. At lower cadence or lower spectral resolution, these components could easily appear blended or be mistaken for a single evolving feature. In our data, however, they are clearly resolved as distinct absorption components. %We suggest that this intermediate component is formed inside a shell produced by interaction between flows of different velocities.  

\begin{figure*}
\begin{center}
\includegraphics[width=1\textwidth]{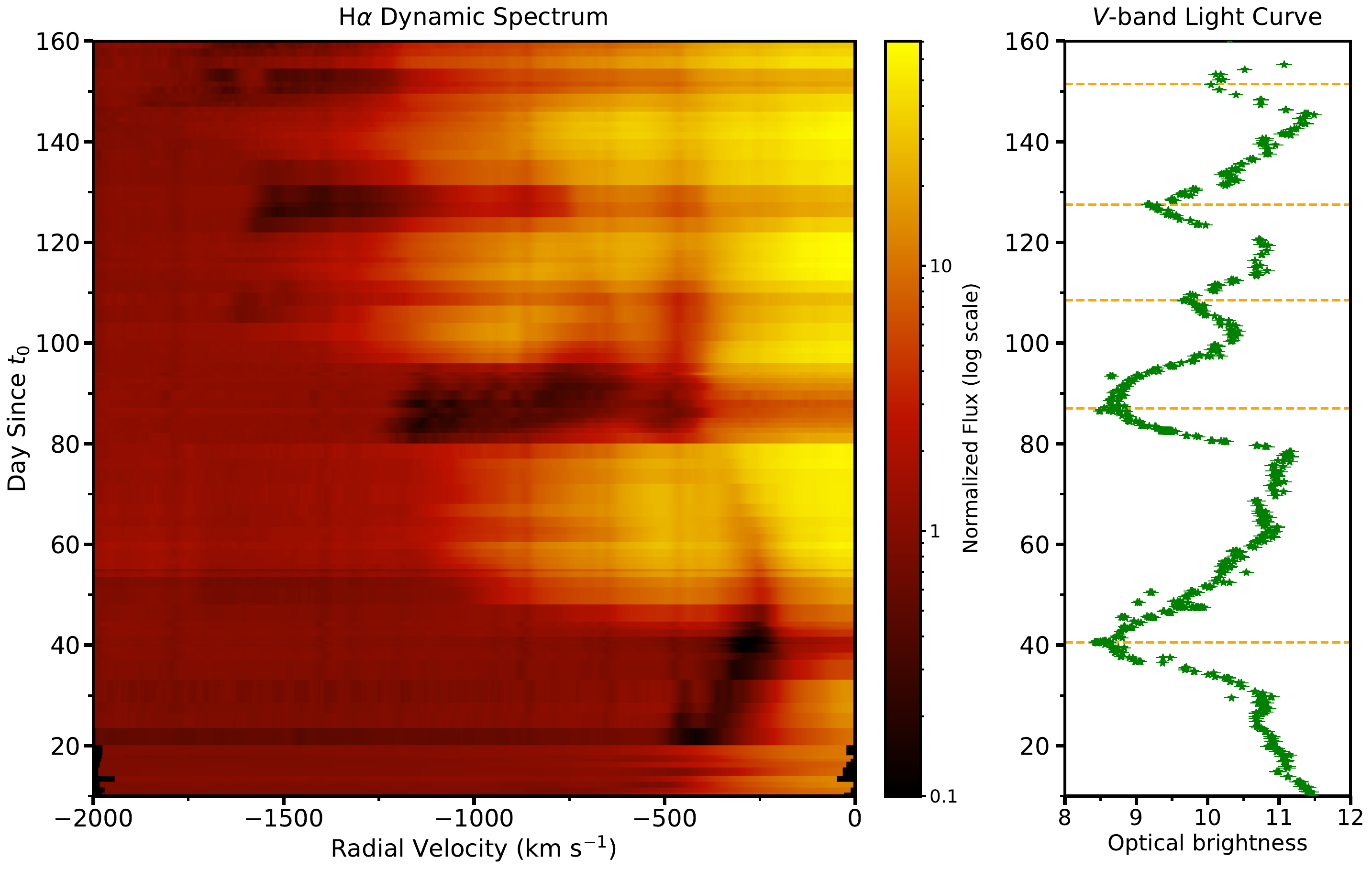}
\caption{\textit{Left}: 2D dynamic spectrum of nova V612~Sct illustrating the evolution of the H$\alpha$ line profiles during the first 160~days of the eruption. Dark features correspond to absorption components. \textit{Right:} the optical $V$-band light curve of V612~Sct during the same time frame. The error bars represent 1-$\sigma$ uncertainties.}
\label{Fig:Dynamic_spec}
\end{center}
\end{figure*}

While most novae typically exhibit two velocity components, a slower and a faster one (and an intermediate one if the cadence is extensive; \citealt{Aydi_etal_2020b}), flaring novae often show a sequence of absorption components appearing at progressively higher velocities, many of which coincide with maxima in the light curve \citep{Tanaka_etal_2011_159,Tanaka_etal_2011,Aydi_etal_2019_I,Aydi_etal_2020a,Aydi_etal_2026}. Although the emergence of these higher-velocity absorption features can be seen in the evolution of the H$\alpha$ line profiles (Figures~\ref{Fig:line_profile_1}--\ref{Fig:line_profile_5}), they are shown more clearly in the two-dimensional dynamic spectrum of H$\alpha$, which shows the spectra of V612~Sct in velocity space, arranged chronologically alongside the light curve (Figure~\ref{Fig:Dynamic_spec}). In this plot, the dark regions correspond to absorption features, which are seen to appear around the peak times of the nova flares and at progressively higher velocities than those associated with earlier flares. Both the two-dimensional dynamic spectrum and the line-profile evolution show that absorption features at different velocities coexist within the same line, in this case H$\alpha$ (for example at features at $\approx -500$, $\approx -1000$, and $\approx -1500$\,km\,s$^{-1}$), implying the presence of multiple ejecta components characterized by distinct velocities.

\subsection{Sonification of the spectral evolution}

To complement the visual analysis of the spectral evolution, we produced a sonification of the H$\alpha$ line profiles throughout the flaring period. The goal is to provide an alternative representation of the time-dependent changes in the line structure, particularly the evolution of the emission profile and the appearance, disappearance, and migration of absorption components in velocity space. In addition to its value as an outreach and communication tool, sonification offers a different way to examine temporal patterns in the data.

The sonification was constructed from the H$\alpha$ line profiles in velocity space. For each spectrum, the wavelength axis was converted to radial velocity, $v$, relative to the rest wavelength of H$\alpha$. The spectra were restricted to the wavelength interval 6510--6670\,\AA, and each epoch was interpolated onto a fixed velocity grid. This ensured that the same velocity interval was always associated with the same sound frequency throughout the sequence.

The fixed velocity grid was then mapped onto audio frequencies between 70 and 600\,Hz using a logarithmic scaling,
\begin{equation}
f(v) = f_{\rm min}\,\exp\left[
\ln\left(\frac{f_{\rm max}}{f_{\rm min}}\right)
\frac{v-v_{\rm min}}{v_{\rm max}-v_{\rm min}}
\right],
\end{equation}
where $f_{\rm min} = 70\,Hz$ and $f_{\rm max}=600\,Hz$ are the minimum and maximum audio frequencies, respectively. A logarithmic mapping was adopted because it provides a smoother perceptual distribution of pitch than a linear mapping.

The flux profile at each epoch was converted into an amplitude envelope after baseline subtraction, normalization, and mild compression. Specifically, a low-level baseline was estimated from the 5th percentile of the flux distribution within the selected spectral window and subtracted from the data. The resulting profile was then normalized to its maximum value and transformed as
\begin{equation}
A(v) \propto \left[\frac{F(v)}{F_{\rm max}}\right]^{\gamma},
\end{equation}
with $\gamma = 0.85$, in order to compress the dynamic range while preserving the overall line morphology. A light smoothing was also applied across adjacent velocity bins.

Rather than assigning each velocity bin to a pure sinusoidal tone, we used narrowband-noise carriers centered on the corresponding audio frequencies. This choice avoids the overly artificial sound produced by large stacks of pure tones, while still preserving the relation between spectral structure and pitch. The synthesized signal can be written schematically as
\begin{equation}
y(t) = \sum_{k=1}^{N} A_k(t)\,\sin[\phi_k(t)],
\end{equation}
where $A_k(t)$ is the amplitude associated with the $k$th velocity bin and $\phi_k(t)$ is the corresponding phase. The phase was evolved continuously in time as
\begin{equation}
\phi_k(t+\Delta t) = \phi_k(t) + 2\pi \frac{f_k + \delta f_k(t)}{f_s}\,\Delta t,
\end{equation}
where $f_k$ is the central frequency of the $k$th bin, $\delta f_k(t)$ is a small stochastic frequency perturbation used to produce narrowband noise, and $f_s$ is the audio sampling rate. This procedure yields a more natural sound while retaining the continuity of the spectral evolution.

To represent the temporal development of the nova continuously, the amplitude envelopes of consecutive spectra were interpolated linearly in time, producing a smooth transition from one epoch to the next without discontinuities in phase. The final audio track therefore encodes the changing structure of the H$\alpha$ line profile throughout the eruption in a continuous manner. The sonification is available online at \footnote{\url{https://www.dropbox.com/scl/fi/vsvp3hw3815n9mpkgd1h1/Sonnification_final_paper.mp4?rlkey=z3wu3565632yugh6g6x6dc3bn&st=f5ao0u6w&dl=0}}.

\section{Discussion and Conclusions}
\label{sec:discussion}
\subsection{Distance to the Nova}

Previous studies provide useful constraints on the distance to V612~Sct. \citet{Schaefer_2022}'s comprehensive nova distance catalog uses {\it Gaia} parallaxes combined with other methods, such as expansion parallaxes and reddening maps, to derive the best available distances for Galactic novae. The {\it Gaia} parallax of V612~Sct is measured as 0.25$\pm$0.75\,mas. In \citet{Schaefer_2022}'s catalog, V612~Sct is assigned a distance of $8063$~pc, with a 68\% confidence range of $7330$--$8811$~pc. More recently,  \citet{Craig_etal_2026} derived distances for novae lacking reliable {\it Gaia} parallaxes using three-dimensional dust maps and a Galactic mass model, obtaining a distance of $9.6 \pm 4.4$~kpc for V612~Sct. Both estimates place the nova in the Galactic bulge region and are consistent within uncertainties. Given the weaker {\it Gaia} parallax for this source and the agreement between the two methods, we adopt a distance of approximately $8$~kpc for V612~Sct in this work.

\begin{figure*}
\begin{center}
\includegraphics[width=0.8\textwidth]{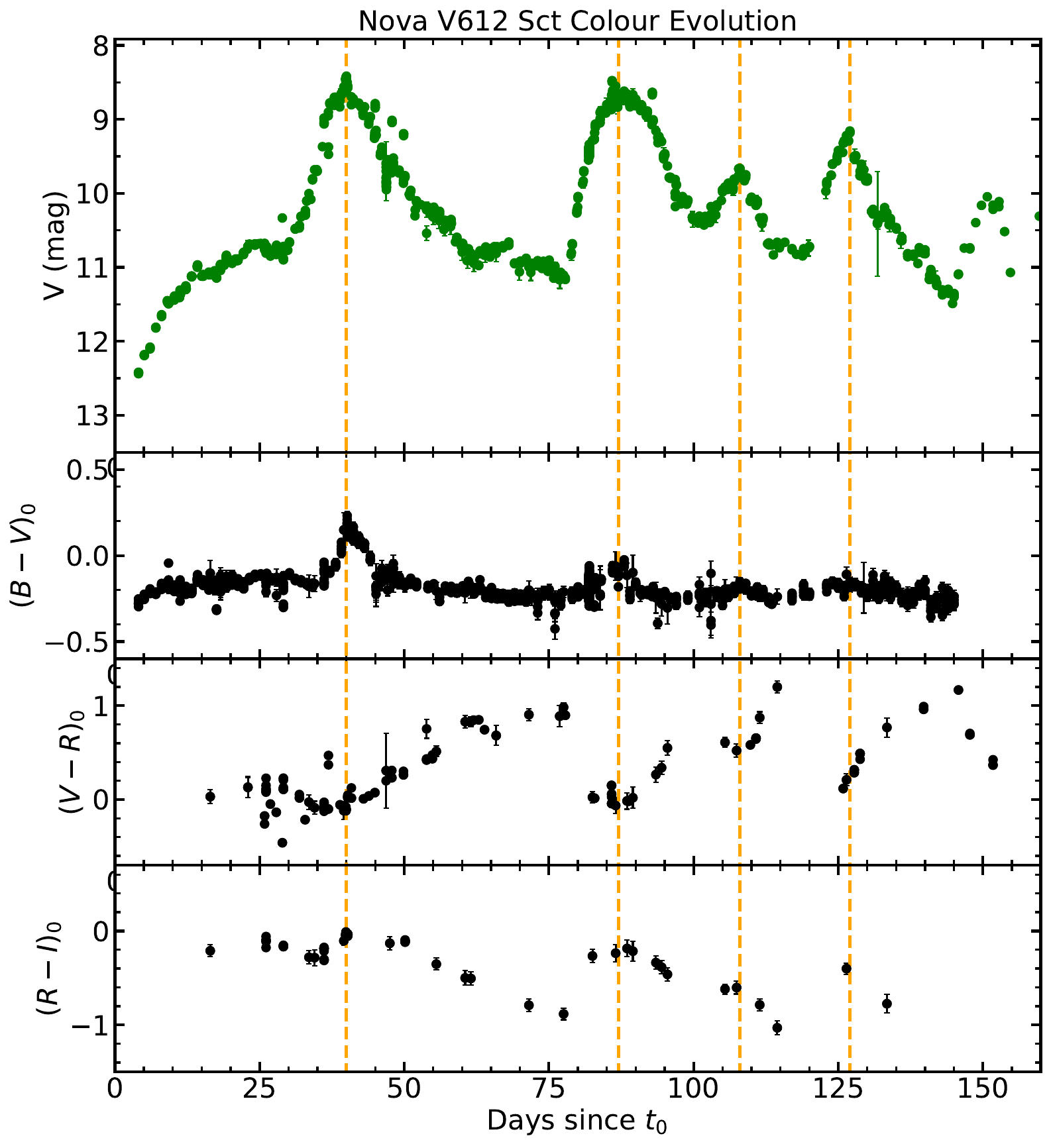}
\caption{From \textit{top} to \textit{bottom}: the $V$-band light curve, and the extinction-corrected colors $(B-V)_0$, $(V-R)_0$, and $(R-I)_0$ of nova V612~Sct. The dashed lines mark the centers of the flares. The colors were computed only for epochs in which observations in the relevant two bands were obtained on the same night. The error bars represent 1-$\sigma$ uncertainties.}
\label{Fig:color_evolution}
\end{center}
\end{figure*}

\subsection{The origin of the flares}
A number of mechanisms have been proposed to explain the flares commonly observed in nova light curves. One widely supported interpretation is that the flares arise from multiple, discrete mass-ejection episodes occurring while nuclear burning is still active on the white dwarf surface \citep{Pejcha_2009,Aydi_etal_2019_I,Aydi_etal_2020a,Aydi_etal_2026}. Each new ejection produces fresh material with different densities and velocities, and the expansion of the material can temporarily increase the optical luminosity, producing an increase in the light curve flux. Moreover, the freshly ejected outflows could interact with previously ejected material, leading to shock formation. The shocks heat up the gas, leading to thermal emission that dissipates through the ejecta and escapes as visible light \citep{Metzger_etal_2014,Metzger_etal_2015,Li_etal_2017_nature,Aydi_etal_2020a}. Other proposed explanations include variations in the luminosity of the white dwarf due to instabilities in the burning envelope, or oscillations in the expanding envelope itself \citep{Chocol_Pribulla_1998,Mason_etal_2020}.

In the case of V612~Sct, our data favor the multiple-ejection scenario. Each flare in the optical light curve coincides with the appearance of new absorption systems at progressively higher velocities in the spectra (Figure~\ref{Fig:Dynamic_spec}). Since these multiple absorption components co-exist within the same lines, they imply the presence of several kinematically distinct shells. These features, together with the evolving ionization state of the ejecta, point toward repeated mass-ejection events and the internal shocks they might generate as the primary drivers of the flaring behavior. Evidence for shock interaction can also be traced in the spectral evolution, particularly in the behavior of the intermediate (or principal) component, which emerges near optical maximum and gradually replaces the slower pre-maximum component. If this component indeed originates in a shell formed through the interaction between the early slow flow and the later faster flow, as proposed by \citet{Friedjung_1987} and \citet{Aydi_etal_2020b}, then it provides additional indirect evidence for shocks developing within the nova ejecta.

Multiple ejections naturally lead to internal shocks when material of different velocities collide. Such shocks are now recognized as a major source of high-energy emission in novae. In several well-studied systems, including V906~Car and V5668~Sgr, the $\gamma$-ray emission detected by \textit{Fermi}-LAT coincided closely with optical flares/maxima, demonstrating that shocks can be responsible for both the high-energy radiation and the optical variability \citep{Cheung_etal_2016,Aydi_etal_2020a,Craig_etal_2026}. These novae provide clear observational evidence that optical flares can arise from shock interaction between distinct ejecta components.

When comparing V612~Sct to these $\gamma$-ray–detected, flaring novae, we find striking similarities in the optical behavior. The light curve of V612~Sct shows multiple flares, and each flare coincides with the appearance of new absorption systems at progressively higher velocities. Although V612~Sct was not detected by \textit{Fermi}-LAT \citep{Craig_etal_2026}, this non-detection is fully consistent with its large distance. As discussed earlier, distance estimates of $\sim$8--10~kpc \citep{Schaefer_2022} place the nova well beyond the range at which typical nova $\gamma$-ray luminosities can be detected by the LAT on \textit{Fermi} \citep{Craig_etal_2026}. The absence of a $\gamma$-ray detection therefore does not argue against the presence of shocks. Instead, the optical and spectroscopic signatures of V612~Sct support a shock-powered scenario, similar to those observed in novae with confirmed $\gamma$-ray emission.

Not only observations, but also the 1D hydrodynamical modeling of \citet{Steinberg_etal_2020} support the idea that flares in novae can arise from collisions between outflows of different velocities. In their picture, the ejecta are not treated as a sequence of strictly discrete shells, but rather as a time-variable outflow that switches between slow and fast modes, naturally generating a train of internal shocks as faster material catches up with slower ejecta. This scenario can reproduce several hallmark properties of flaring novae, including correlated optical and $\gamma$-ray flares, temporary photospheric expansion, and complex spectral-line evolution with accelerating, decelerating, and merging velocity components. From this perspective, the presence of multiple ejecta components at different velocities in V612~Sct is fully consistent with a shock-powered flare scenario in which repeated changes in the outflow velocity structure drive successive collisions within the nova ejecta.

We also examine the color evolution during the flaring phase (Figure~\ref{Fig:color_evolution}). As observed in other flaring novae, the flares are accompanied by clear changes in the broadband colors. Such variations are often interpreted as evidence for changes in the radius, and hence the temperature, of the emitting photosphere. However, in novae, the broadband colors can also be strongly influenced by emission lines, since the spectra are often dominated by line emission. Therefore, changes in the line strengths relative to the continuum across different bandpasses can produce substantial color variations. This is particularly true in the $B$, $V$, and $R$, bands where strong Balmer, Fe~II, He~I, and O~I lines exist, considerably affecting the $(B-V)_0$ and $(V-R)_0$ colors. Given the clear evolution of the emission-line strengths relative to the continuum shown in Figures~\ref{Fig:line_profile_1}--\ref{Fig:line_profile_5} and Figures~\ref{Fig:spec_1}--\ref{Fig:spec_13}, we suggest that line variability contributes significantly to the observed color changes. At the same time, the evolution of the line species and ionization state during the flaring phase strongly indicates that the temperature of the emitting region (the photosphere), is also changing. This evolution may be driven by both changes in the location of the emitting region and the development of shocks within the ejecta.

\begin{figure*}
\begin{center}
\includegraphics[width=0.8\textwidth]{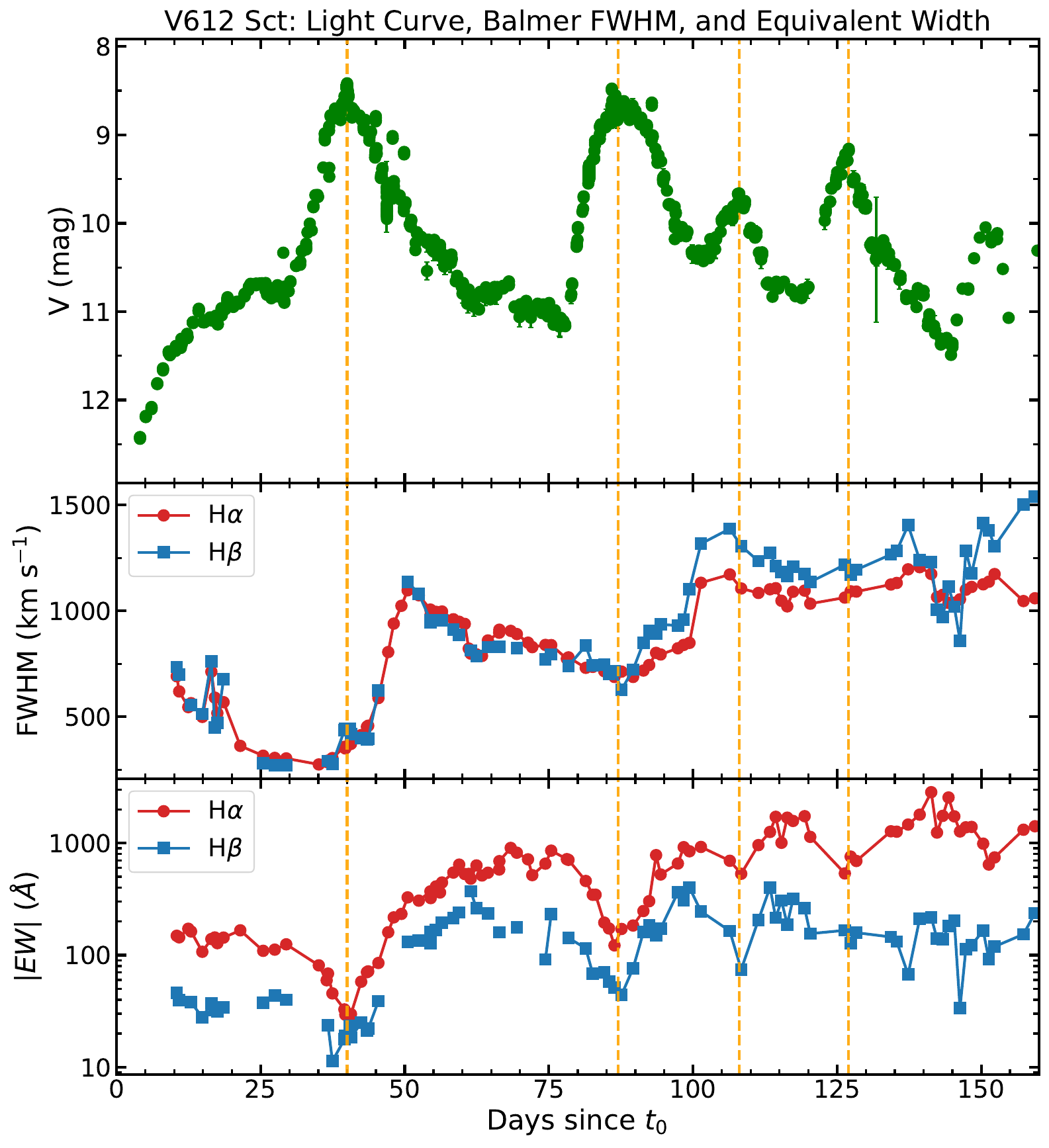}
\caption{From \textit{top} to \textit{bottom}: the $V$-band light curve, the full width at half maximum (FWHM), and the absolute equivalent width, $|\mathrm{EW}|$, of H$\alpha$ (red) and H$\beta$ (blue) for the nova V612~Sct. The vertical dashed lines mark the centers of the flares. Since emission lines have negative EWs by convention, their absolute values are shown here.}
\label{Fig:LC_FWHM_EW}
\end{center}
\end{figure*}

Figure~\ref{Fig:LC_FWHM_EW} shows the evolution of the full width at half maximum (FWHM) and equivalent width (EW) of the Balmer lines in relation to the optical light curve. Both the measured line widths and strengths generally decrease during the flares and increase again during the intervals between successive flares. The decrease in EW is naturally explained by the substantial brightening of the continuum relative to the emission-line flux during each flare (see also Figure~\ref{Fig:He_LC}). The accompanying decrease in FWHM could reflect changes in the line-forming region or ejecta kinematics. However, it may also arise partly from the changing line-to-continuum contrast: as the continuum brightens, the broad, low-intensity wings become less prominent, and the measured profile is increasingly dominated by the narrower central peak. Between flares, the Balmer lines (along with other line species) strengthen relative to the continuum, and their changing contributions to the broadband fluxes can therefore produce substantial variations in the observed color indices.

\subsection{Is V612~Sct a nova impostor?}
\citet{Mason_etal_2020} suggested that slow, flaring novae such as V612~Sct may represent a distinct class of transient phenomena, which they termed \emph{nova impostors}. They argued that, although the observed properties of V612~Sct are broadly consistent with those of a nova eruption, the presence of major optical flares may point to a phenomenon that resembles a nova without necessarily being a standard \textit{classical nova}. Novae do indeed display a remarkable diversity of behavior, particularly at optical wavelengths: some evolve rapidly and smoothly, whereas others exhibit unusual light-curve morphologies, including long-lasting plateaus near maximum, multiple peaks, oscillations, and cusp-shaped structures \citep{Strope_etal_2010}. The physical origin of this diversity is still not fully understood. One may therefore ask whether such behavior reflects a diversity of underlying phenomena rather than a single class grouped under the term ``novae.''

While this remains a possibility, it is also plausible that the observed diversity arises naturally from the wide range of physical parameters that characterize nova systems (see, e.g., \citealt{Yaron_etal_2005,Hillman_etal_2014,Hachisu_Kato_2015}). The white dwarf mass, the mass-accretion rate, the accumulated envelope mass at ignition, the chemical composition of the white dwarf, and the binary separation may all play important roles in shaping the observational properties of the eruption. In addition, the increasingly recognized role of shocks in powering part of the optical emission in novae (e.g., \citealt{Li_etal_2017_nature,Aydi_etal_2020a}) may also help explain some of the more unusual features seen in nova light curves.

In the case of V612~Sct, the white dwarf is suggested to be on the very low end of white dwarf masses in nova systems ($M \sim 0.4\,M_{\odot}$; \citealt{Mason_etal_2020}), implying that a relatively massive accreted envelope would be required to trigger the eruption \citep{Shara_1981,Livio_1992,Shara_etal_1993}. In such a system, once the thermonuclear runaway begins, the bulk of the envelope may engulf the binary but not be expelled instantaneously, as discussed above. This could naturally produce a slow rise toward optical maximum (see \citealt{Aydi_etal_2026}). Even after most of the envelope is ejected, the low outflow velocities, of only a few hundred km\,s$^{-1}$, raise the possibility that part of the material may remain only marginally unbound, or even fall back toward the white dwarf, leading to renewed accretion and possibly subsequent ejection episodes. \citet{Mason_etal_2020} found that the ejected mass in V612~Sct ($7\times10^{-4} \lesssim M_{\rm ej} \lesssim 9\times10^{-4}\,M_\odot$) is about an order of magnitude larger than the ejecta masses typically inferred for classical novae \citep{Yaron_etal_2005}. This supports the possibility that the prolonged flaring phase is associated with multiple, physically distinct ejection episodes rather than a single, impulsive mass-loss event. In this picture, the flares would not require V612~Sct to belong to a fundamentally different class of transients, but could instead reflect an extreme manifestation of nova behavior in a system with an unusually low-mass white dwarf, a massive envelope, and extended mass ejection that could lead to shock interaction.

\subsection{Sonification as a scientific and accessibility tool}
\label{sec:discussion_sonification}

The sonification presented in this work provides an alternative representation of the spectroscopic evolution of V612~Sct and highlights the potential utility of auditory analysis in time-domain astrophysics. By converting the evolving spectral line profiles into sound, changes in the velocity structure and the appearance of new absorption systems become perceptible through variations in pitch, intensity, and temporal structure. In particular, the emergence of progressively higher-velocity absorption components during successive flares produces distinct auditory signatures that qualitatively trace the possible repeated mass-ejection episodes inferred from the optical spectra.

While sonification is often presented primarily as an outreach or educational technique, it may also offer scientific advantages when applied to large and complex spectroscopic datasets. Human auditory perception is highly sensitive to temporal and frequency variations, making it potentially useful for identifying subtle patterns, evolving structures, or anomalies that may be difficult to recognize visually in static spectra or even dynamic spectral maps. In the case of V612~Sct, the repeated appearance of new blue-shifted absorption troughs during the flaring episodes creates recognizable changes in the soundscape that parallel the spectroscopic evidence for possible multiple outflows. Such approaches may become increasingly valuable in the era of time-domain surveys, where rapidly evolving multi-epoch datasets require complementary methods of exploration and pattern recognition.

Sonification may also prove useful for comparative studies of novae and other transients. Different classes of eruptions, spectral morphologies, or ejecta geometries may produce distinct auditory characteristics, potentially enabling rapid qualitative comparisons between events. In systems where line blending, complex velocity structure, or low signal-to-noise ratios complicate visual inspection, auditory representations could provide an additional avenue for identifying coherent structures or transient features worthy of further quantitative analysis.

Beyond its scientific applications, sonification contributes to improving the accessibility of astronomical data for blind and low-vision researchers and members of the public. Astronomy is traditionally a highly visual field, and converting data into auditory form provides an alternative means of engaging with spectroscopic and time-domain phenomena. Developing accessible analysis tools broadens participation in astrophysical research while also encouraging new perspectives and methodologies for interacting with complex datasets. Although sonification is unlikely to replace traditional visualization or quantitative spectral analysis, it can serve as a complementary tool that combines scientific exploration, data accessibility, and public engagement. Future work applying sonification systematically to larger samples of novae and other transients may help determine its effectiveness for identifying recurring spectroscopic behaviors, shock signatures, and multi-phase outflows.
 
\section{Conclusions}

We have presented a comprehensive photometric and spectroscopic study of the slow, highly flaring nova V612~Sct, using public AAVSO photometry and densely sampled ARAS spectroscopy. The nova exhibits an unusually slow evolution, with $t_2 \gtrsim 114$~days, and a complex optical light curve characterized by at least five maxima during the first $\sim$160 days of the eruption. The first two flares are particularly prominent, with amplitudes of $\sim$2.4--2.5 mag and durations of several weeks, while the later flares are shorter and less regular. The spacing between successive maxima does not follow the logarithmic-time trend reported for some other flaring novae, but instead shows an overall decrease with time.

The spectroscopic evolution reveals repeated and systematic changes that are tightly linked to the flaring behavior. V612~Sct evolves from an early He/N phase into a Fe~II phase, and later alternates repeatedly between He/N and Fe~II characteristics in step with the optical flares. This behavior indicates substantial and recurrent changes in the ionization state and optical depth of the ejecta during the eruption. The color evolution during the flaring phase is similarly complex, and is likely driven by a combination of changes in the photospheric temperature and radius, together with strong variations in the relative contributions of emission lines and continuum flux within the broadband filters.

The H$\alpha$ line-profile evolution provides particularly strong evidence for multiple phases of mass loss. During the rise to maximum, the nova shows a low-velocity pre-maximum absorption component that gradually shifts to slower apparent velocities, consistent with the recession of the photosphere through an expanding outflow. After maximum light, broader emission wings and additional absorption systems appear, demonstrating the coexistence of multiple kinematic components within the ejecta. In particular, we identify a slow component, a faster component, and an intermediate or principal component that emerges near optical maximum and persists through the subsequent evolution. The dynamic spectrum further shows that new absorption systems appear at progressively higher velocities around the times of the flares, suggesting repeated ejection episodes.

Taken together, the photometric and spectroscopic data favor a scenario in which the flares are powered by multiple, discrete mass-ejection episodes and the internal shocks produced when faster material overtakes slower ejecta. The emergence of progressively faster absorption systems, the coexistence of multiple velocity components in the same lines, and the appearance of the intermediate/principal component all support this interpretation. Although V612~Sct was not detected in $\gamma$-rays, its large distance likely explains this non-detection, and the optical and spectroscopic signatures closely resemble those of other shock-powered flaring novae.

Our results do not require V612~Sct to represent a fundamentally different class of transient. Instead, the nova can be understood as an extreme but physically plausible member of the classical nova population. Its unusually low inferred white dwarf mass, massive accreted envelope, slow ejecta velocities, and high total ejected mass may naturally produce an extended eruption in which mass loss occurs over multiple episodes rather than in a single impulsive event. In this picture, the prolonged flaring activity reflects the combined effects of repeated ejection and shock interaction within the ejecta.

Finally, we have complemented the visual analysis with a sonification of the H$\alpha$ line-profile evolution, providing an alternative way to follow the temporal development of the ejecta structure. While primarily intended as a complementary and outreach-oriented representation of the data, it also illustrates the remarkable complexity and continuity of the spectral evolution during the flaring phase.

Future multiwavelength observations of similarly slow, flaring novae, especially coordinated optical, X-ray, and $\gamma$-ray monitoring with dense time-resolved spectroscopy, will be essential for determining how common such repeated ejection and shock-powered evolution may be, and for clarifying the physical origin of the most extreme forms of nova variability.

\section*{Acknowledgments}

We thank the ARAS and AAVSO observers from around the world who contributed their spectra to the ARAS database and their magnitude measurements to the AAVSO International Database, used in this work.
Nova astrophysics at Michigan State University (MSU) is supported by NASA grants 80NSSC23K0497 and 80NSSC25K7334 and the MSU Jenison Fund.

BDM acknowledges support from NASA (grants 80NSSC22K0807, 80NSSC24K0408, and 80NSSC26K0300).  The Flatiron Institute is supported by the Simons Foundation. 

SM acknowledges support from NASA (TM3-24002A) and the Virginia Institute for Theoretical Astrophysics (VITA), supported by the College and Graduate School of Arts and Sciences at the University of Virginia.

Analysis made significant use of \textsc{python} 3.7.4, and the associated packages \textsc{numpy}, \textsc{matplotlib}, \textsc{seaborn}, \textsc{scipy}. 

\section*{Data availability}
The AAVSO and ASAS-SN data used in this work are publicly available. Same for the ARAS spectra, nevertheless a copy of the spectra used can be found here: \url{https://www.dropbox.com/scl/fi/c6t6xrawnkvtf3qx4cuvm/V612_spectra_by_day_halpha.tar.xz?rlkey=yhrag19267gat0afrqaota2tg&st=ksw4i3w1&dl=0}.\\
The supplementary plots of this paper are included in the Appendix. 
%\bibliographystyle{apalike}
%\bibliography{references}

%\bibliographystyle{mnras_vanHack}
\bibliography{biblio}

\appendix

\renewcommand\thetable{\thesection.\arabic{table}}    
\renewcommand\thefigure{\thesection.\arabic{figure}}   
\setcounter{figure}{0}

\section{Supplementary plots and tables}
\label{appB}
In this Appendix we present supplementary plots and tables.

\begin{table*}
\centering
\caption{Spectroscopic observation logs of nova V612~Sct.}
\begin{tabular}{cccccc}
\hline
Source & date & $ t - t_0$ & Resolving power & $\lambda$ Range & Observer \\ 
 & &(days) &  & (\AA)\\
\hline
ARAS & 2017-06-29T20:24:45 & 10.45 & 580 & 4377\,--\,7355 & Paolo Berardi\\
ARAS & 2017-06-30T06:29:31 & 10.87 & 622 & 3866\,--\,7448 & Woody Sims\\
ARAS & 2017-07-01T20:29:25 & 12.45 & 2650 & 6056\,--\,7059 & Paolo Berardi \\
ARAS & 2017-07-02T07:59:18 & 12.93 & 664 & 3867\,--\,7449 & Woody Sims\\
ARAS & 2017-07-04T06:43:18 & 14.88 & 966 & 3870\,--\,7454 & Woody Sims\\
ARAS & 2017-07-05T20:30:58 & 16.45 & 579 & 3950\,--\,7360 & Umberto Sollecchia\\
ARAS & 2017-07-06T10:54:33 & 17.05 & 1491 & 3800\,--\,7267 & TBohlsen\\
ARAS & 2017-07-06T21:21:46 & 17.49 & 540 & 3846\,--\,7375 & cbuil\\
ARAS & 2017-07-07T23:19:03 & 18.57 & 669 & 3975\,--\,7546 & Olivier Garde\\
ARAS & 2017-07-10T20:53:29 & 21.47 & 5491 & 6268\,--\,6731 & Paolo Berardi\\
ARAS & 2017-07-14T20:15:19 & 25.44 & 11000 & 4186\,--\,7314 & Olivier Garde\\
ARAS & 2017-07-16T20:20:33 & 27.45 & 11000 & 4186\,--\,7314 & Olivier Garde\\
ARAS & 2017-07-18T20:12:15 & 29.44 & 11000 & 4298\,--\,7308 & Olivier Garde\\
ARAS & 2017-07-23T12:21:42 & 34.12 & 14846 & 6491\,--\,6660 & TBohlsen\\
ARAS & 2017-07-24T11:39:52 & 35.09 & 14802 & 6491\,--\,6660 & TBohlsen\\
ARAS & 2017-07-25T20:25:43 & 36.45 & 5616 & 6255\,--\,6724 & Paolo Berardi\\
ARAS & 2017-07-26T02:04:16 & 36.69 & 13000 & 4030\,--\,7950 & tlester\\
ARAS & 2017-07-26T20:20:00 & 37.45 & 9000 & 4052\,--\,7498 & J. Guarro\\
ARAS & 2017-07-28T22:16:55 & 39.53 & 11000 & 4186\,--\,7314 & Olivier Garde\\
ARAS & 2017-07-29T02:04:46 & 39.69 & 12000 & 4030\,--\,7950 & tlester\\
ARAS & 2017-07-29T21:38:48 & 40.50 & 9000 & 4052\,--\,7760 & JGF\\
ARAS & 2017-07-30T01:43:45 & 40.67 & 12000 & 4030\,--\,7950 & tlester\\
ARAS & 2017-07-31T20:11:50 & 42.44 & 9000 & 4052\,--\,7498 & J. Guarro\\
ARAS & 2017-08-01T20:12:35 & 43.44 & 9000 & 4052\,--\,7760 & JGF\\
ARAS & 2017-08-02T01:27:27 & 43.66 & 13000 & 4030\,--\,7950 & tlester\\
ARAS & 2017-08-02T12:59:26 & 44.14 & 14212 & 6501\,--\,6613 & Paul Luckas\\
ARAS & 2017-08-03T20:01:38 & 45.43 & 9000 & 4052\,--\,7760 & JGF\\
ARAS & 2017-08-05T12:43:02 & 47.13 & 13926 & 6501\,--\,6613 & Paul Luckas\\
ARAS & 2017-08-06T11:42:55 & 48.09 & 14886 & 6491\,--\,6660 & TBohlsen\\
ARAS & 2017-08-07T19:39:28 & 49.42 & 5702 & 6263\,--\,6729 & Paolo Berardi\\
ARAS & 2017-08-08T21:34:38 & 50.50 & 9000 & 4052\,--\,7760 & JGF\\
ARAS & 2017-08-10T20:08:26 & 52.44 & 9000 & 4052\,--\,7498 & J. Guarro\\
ARAS & 2017-08-12T20:28:23 & 54.45 & 9000 & 4052\,--\,7760 & JGF\\
ARAS & 2017-08-12T21:35:08 & 54.50 & 11000 & 4250\,--\,7306 & Olivier Garde\\
ARAS & 2017-08-13T19:43:31 & 55.42 & 9000 & 4052\,--\,7760 & JGF\\
ARAS & 2017-08-14T12:31:14 & 56.12 & 14762 & 6492\,--\,6660 & TBohlsen\\
ARAS & 2017-08-14T20:44:09 & 56.46 & 9000 & 4052\,--\,7498 & J. Guarro\\
ARAS & 2017-08-16T19:53:53 & 58.43 & 9000 & 4052\,--\,7498 & J. Guarro\\
ARAS & 2017-08-17T20:06:18 & 59.44 & 9000 & 4052\,--\,7498 & J. Guarro\\
ARAS & 2017-08-18T19:22:07 & 60.41 & 16949 & 6479\,--\,6635 & Paolo Berardi\\
ARAS & 2017-08-19T12:03:09 & 61.10 & 14740 & 6492\,--\,6660 & TBohlsen\\
ARAS & 2017-08-19T20:15:14 & 61.44 & 9000 & 4052\,--\,7760 & JGF\\
ARAS & 2017-08-20T20:08:26 & 62.44 & 9000 & 4128\,--\,7760 & JGF\\
ARAS & 2017-08-21T19:08:34 & 63.40 & 16365 & 6481\,--\,6637 & Paolo Berardi\\
ARAS & 2017-08-22T20:02:02 & 64.43 & 9000 & 4052\,--\,7498 & J. Guarro\\
ARAS & 2017-08-24T18:44:51 & 66.38 & 15432 & 6480\,--\,6636 & Paolo Berardi\\
ARAS & 2017-08-24T19:32:01 & 66.41 & 9000 & 4052\,--\,7760 & JGF\\
ARAS & 2017-08-26T18:41:39 & 68.38 & 16685 & 6479\,--\,6636 & Paolo Berardi \\
ARAS & 2017-08-27T20:03:03 & 69.44 & 9000 & 4052\,--\,7498 & J. Guarro\\
ARAS & 2017-08-29T18:44:29 & 71.38 & 16694 & 6479\,--\,6636 & Paolo Berardi\\
ARAS & 2017-08-30T12:33:03 & 72.12 & 14646 & 6493\,--\,6661 & TBohlsen\\
ARAS & 2017-09-01T19:13:07 & 74.40 & 9000 & 4052\,--\,7498 & J. Guarro\\
ARAS & 2017-09-02T19:16:01 & 75.40 & 9000 & 4052\,--\,7760 & JGF\\
ARAS & 2017-09-05T12:27:09 & 78.12 & 14748 & 6493\,--\,6661 & TBohlsen\\
ARAS & 2017-09-05T19:06:14 & 78.40 & 9000 & 4052\,--\,7498 &  J. Guarro\\
ARAS & 2017-09-08T19:08:15 & 81.40 & 9000 & 4052\,--\,7498 & J. Guarro\\
ARAS & 2017-09-10T00:19:29 & 82.61 & 12000 & 4031\,--\,7949 & tlester\\
ARAS & 2017-09-10T12:04:45 & 83.10 & 14733 & 6492\,--\,6663 & TBohlsen\\
ARAS & 2017-09-11T00:16:18 & 83.61 & 13000 & 4030\,--\,7948 & tlester\\
ARAS & 2017-09-12T00:20:25 & 84.61 & 13000 & 4030\,--\,7948 & tlester\\
ARAS & 2017-09-12T19:56:11 & 85.43 & 9000 & 4052\,--\,7760 & JGF\\
ARAS & 2017-09-13T18:47:18 & 86.38 & 9000 & 4052\,--\,7498 & J. Guarro\\
ARAS & 2017-09-14T12:23:26 & 87.12 & 13690 & 6492\,--\,6662 & TBohlsen\\
ARAS & 2017-09-15T00:04:35 & 87.60 & 13000 & 4030\,--\,7948 & tlester\\
\hline
\end{tabular}
\label{table:spec_log_V612_Sct_1}
\end{table*}

\begin{table*}
\centering
\caption{Spectroscopic observation logs of nova V612~Sct (continued).}
\begin{tabular}{cccccc}
\hline
Source & date & $ t - t_0$ & Resolving power & $\lambda$ Range \\ 
 & &(days) &  & (\AA) & Observer Name\\
\hline
ARAS & 2017-09-16T11:33:24 & 89.08 & 14746 & 6492\,--\,6663 & TBohlsen\\
ARAS & 2017-09-17T00:08:07 & 89.61 & 13000 & 4030\,--\,7948 & tlester\\
ARAS & 2017-09-18T19:10:55 & 91.40 & 9000 & 4128\,--\,7760 & JGF\\
ARAS & 2017-09-19T19:08:27 & 92.40 & 9000 & 4052\,--\,7498 & J. Guarro\\
ARAS & 2017-09-20T18:37:39 & 93.38 & 9000 & 4052\,--\,7498 & J. Guarro\\
ARAS & 2017-09-20T23:53:07 & 93.60 & 13000 & 4031\,--\,7948 & tlester\\
ARAS & 2017-09-21T18:54:03 & 94.39 & 9000 & 4052\,--\,7498 & J. Guarro\\
ARAS & 2017-09-22T18:32:02 & 95.37 & 7285 & 6414\,--\,6797 & Umberto Sollecchia\\
ARAS & 2017-09-24T18:47:30 & 97.38 & 9000 & 4052\,--\,7760 & JGF\\
ARAS & 2017-09-25T18:59:56 & 98.39 & 9000 & 4052\,--\,7760 & JGF\\
ARAS & 2017-09-26T18:54:49 & 99.39 & 9000 & 4052\,--\,7760 & JGF\\
ARAS & 2017-09-28T18:38:47 & 101.38 & 9000 & 4052\,--\,7498 & J. Guarro\\
ARAS & 2017-09-30T18:19:38 & 103.36 & 7080 & 6374\,--\,6750 & Umberto Sollechia\\
ARAS & 2017-10-02T18:14:07 & 105.36 & 8203 & 6368\,--\,6753 & Umberto Sollechia\\
ARAS & 2017-10-03T09:51:29 & 106.01 & 13610 & 6493\,--\,6660 & TBohlsen\\
ARAS & 2017-10-03T18:51:13 & 106.39 & 9000 & 4052\,--\,7498 & J. Guarro\\
ARAS & 2017-10-04T18:32:38 & 107.37 & 8522 & 6368\,--\,6757 & Umberto Sollechia\\
ARAS & 2017-10-05T09:57:22 & 108.01 & 14495 & 6493\,--\,6660  & TBohlsen\\
ARAS & 2017-10-05T18:07:56 & 108.36 & 9000 & 4052\,--\,7498 & J. Guarro\\
ARAS & 2017-10-08T18:07:05 & 111.35 & 9000 & 4130\,--\,7757 & JGF\\
ARAS & 2017-10-09T09:53:48 & 112.01 & 14298 & 6493\,--\,6659 & TBohlsen\\
ARAS & 2017-10-10T18:21:31 & 113.36 & 9000 & 4052\,--\,7498 & J. Guarro\\
ARAS & 2017-10-11T17:58:07 & 114.35 & 9000 & 4052\,--\,7498 & J. Guarro\\
ARAS & 2017-10-12T10:46:48 & 115.05 & 14240 & 6493\,--\,6659 & TBohlsen\\
ARAS & 2017-10-12T17:37:34 & 115.33 & 8874 & 6371\,--\,6772 & Umberto Sollechia\\
ARAS & 2017-10-12T18:02:06 & 115.35 & 9000 & 4128\,--\,7760 & JGF\\
ARAS & 2017-10-13T17:57:42 & 116.35 & 9000 & 4052\,--\,7498 & J. Guarro\\
ARAS & 2017-10-14T18:00:20 & 117.35 & 9000 & 4052\,--\,7760 & JGF\\
ARAS & 2017-10-16T18:23:39 & 119.37 & 9000 & 4128\,--\,7760 & JGF\\
ARAS & 2017-10-17T17:48:56 & 120.34 & 9000 & 4052\,--\,7498 & J. Guarro\\
ARAS & 2017-10-19T17:22:08 & 122.32 & 8339 & 6389\,--\,6759 & Umberto Sollechia\\
ARAS & 2017-10-20T17:52:44 & 123.35 & 9000 & 4052\,--\,7760 & JGF\\
ARAS & 2017-10-21T10:10:34 & 124.02 & 14976 & 6493\,--\,6659 & TBohlsen\\
ARAS & 2017-10-23T17:21:46 & 126.32 & 8998 & 6387\,--\,6763 & Umberto Sollecchia\\
ARAS & 2017-10-23T17:45:32 & 126.34 & 9000 & 4052\,--\,7760 & JGF\\
ARAS & 2017-10-24T09:44:27 & 127.01 & 14547 & 6493\,--\,6659 & TBohlsen\\
ARAS & 2017-10-24T17:54:06 & 127.35 & 9000 & 4052\,--\,7498 & J. Guarro\\
ARAS & 2017-10-25T17:41:12 & 128.34 & 9000 & 4052\,--\,7498 & J. Guarro\\
ARAS & 2017-10-31T09:34:39 & 134.00 & 14490 & 6493\,--\,6659 & TBohlsen\\
ARAS & 2017-10-31T17:32:25 & 134.33 & 9000 & 4052\,--\,7760 & JGF\\
ARAS & 2017-11-01T17:43:39 & 135.34 & 9000 & 4052\,--\,7760 & JGF\\
ARAS & 2017-11-03T17:26:57 & 137.32 & 9000 & 4052\,--\,7498 & J. Guarro\\
ARAS & 2017-11-05T17:34:42 & 139.33 & 9000 & 4052\,--\,7498 & J. Guarro\\
ARAS & 2017-11-07T17:36:00 & 141.33 & 9000 & 4052\,--\,7760 & JGF\\
ARAS & 2017-11-08T17:37:50 & 142.33 & 9000 & 4052\,--\,7760 & JGF\\
ARAS & 2017-11-09T17:26:38 & 143.33 & 9000 & 4052\,--\,7498 & J. Guarro\\
ARAS & 2017-11-10T17:29:23 & 144.33 & 9000 & 4052\,--\,7760 & JGF\\
ARAS & 2017-11-11T17:28:52 & 145.33 & 9000 & 3980\,--\,7498 & J. Guarro\\
ARAS & 2017-11-12T17:20:34 & 146.32 & 9000 & 3980\,--\,7498 & J. Guarro\\
ARAS & 2017-11-13T17:26:17 & 147.33 & 9000 & 4052\,--\,7760 & JGF\\
ARAS & 2017-11-14T17:26:46 & 148.33 & 9000 & 4052\,--\,7760 & JGF\\
ARAS & 2017-11-16T17:41:02 & 150.34 & 9000 & 4052\,--\,7760 & JGF\\
ARAS & 2017-11-17T17:19:37 & 151.32 & 9000 & 4052\,--\,7760 & JGF\\
ARAS & 2017-11-18T17:22:29 & 152.32 & 9000 & 4052\,--\,7760 & JGF\\
ARAS & 2017-11-22T17:19:55 & 156.32 & 9000 & 4052\,--\,7498 & J. Guarro\\
ARAS & 2017-11-23T17:35:10 & 157.33 & 9000 & 4244\,--\,7495 & J. Guarro\\
ARAS & 2017-11-25T17:23:18 & 159.32 & 9000 & 4052\,--\,7760 & JGF\\
ARAS & 2017-11-26T17:20:29 & 160.32 & 9000 & 3980\,--\,7498 & JGF\\
ARAS & 2017-12-01T16:39:36 & 165.29 & 2547 & 5954\,--\,7334 & Umberto Sollecchia\\
ARAS & 2017-12-06T16:37:47 & 170.29 & 2387 & 5884\,--\,7288 & Umberto Sollecchia\\
ARAS & 2018-04-23T01:36:32 & 307.67 & 7361 & 6354\,--\,6752 & Umberto Sollecchia\\
ARAS & 2018-07-17T12:02:38 & 393.10 & 1694 & 3800\,--\,7380 & TBohlsen\\
\hline
\end{tabular}
\label{table:spec_log_V612_Sct_2}
\end{table*}

\begin{table*}[ht]
\centering
\caption{Sample of AAVSO photometric data used in this study.}
\begin{tabular}{cccc}
\hline
\hline
JD (2450000+) & Filter & Magnitude & Uncertainty \\
\hline
9020.45 & V & 12.84 & 0.03 \\
9021.12 & V & 12.67 & 0.02 \\
9021.88 & V & 12.51 & 0.03 \\
9022.54 & B & 13.42 & 0.04 \\
9022.90 & V & 12.39 & 0.02 \\
9023.33 & R & 12.05 & 0.03 \\
9023.77 & V & 12.28 & 0.02 \\
9024.41 & I & 11.92 & 0.04 \\
9024.95 & V & 12.21 & 0.02 \\
9025.60 & B & 13.28 & 0.05 \\
9026.14 & V & 12.18 & 0.02 \\
9026.82 & R & 11.97 & 0.03 \\
9027.48 & V & 12.16 & 0.02 \\
9028.03 & I & 11.89 & 0.04 \\
9028.71 & V & 12.15 & 0.02 \\
\hline
\end{tabular}
\label{tab:photometry_sample}
\end{table*}

\begin{figure*}
\begin{center}
  \includegraphics[width=\textwidth]{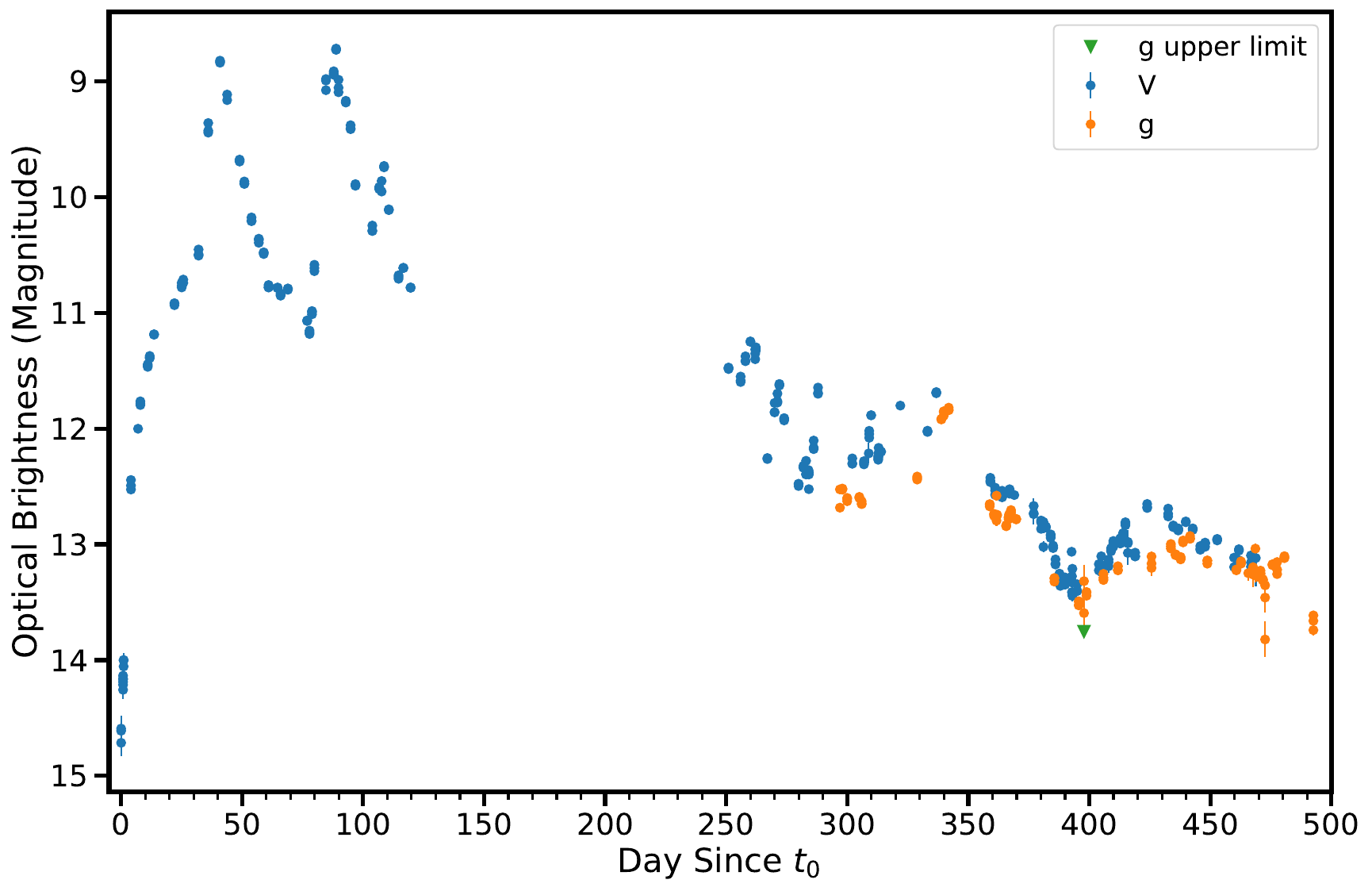}
\caption{The ASAS-SN light curve of V612 Sct over the first 500 days of the eruption. The light curve reveals renewed brightenings of approximately 0.5--1.0\,mag. around days 330 and 430 after $t_0$.} 
\label{Fig:ASASSN_LC}
\end{center}
\end{figure*}

\begin{figure*}
\begin{center}
  \includegraphics[width=0.8\textwidth]{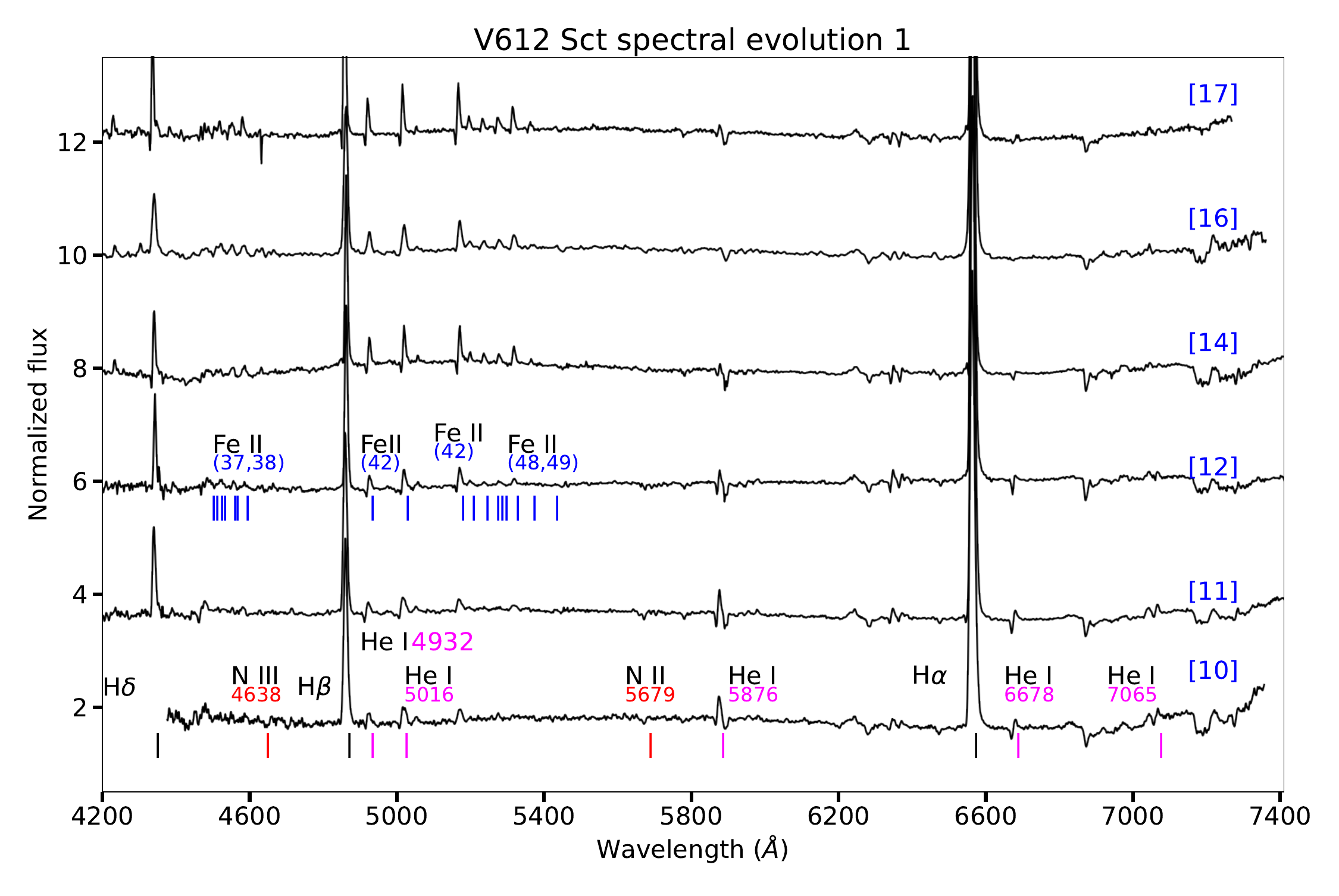}
\caption{The chronological spectral evolution of V612~Sct. The numbers in brackets are days since $t_0$. Colored line identifications are included to help distinguish the spectral features.} 
\label{Fig:spec_1}
\end{center}
\end{figure*}

\begin{figure*}
\begin{center}
  \includegraphics[width=0.8\textwidth]{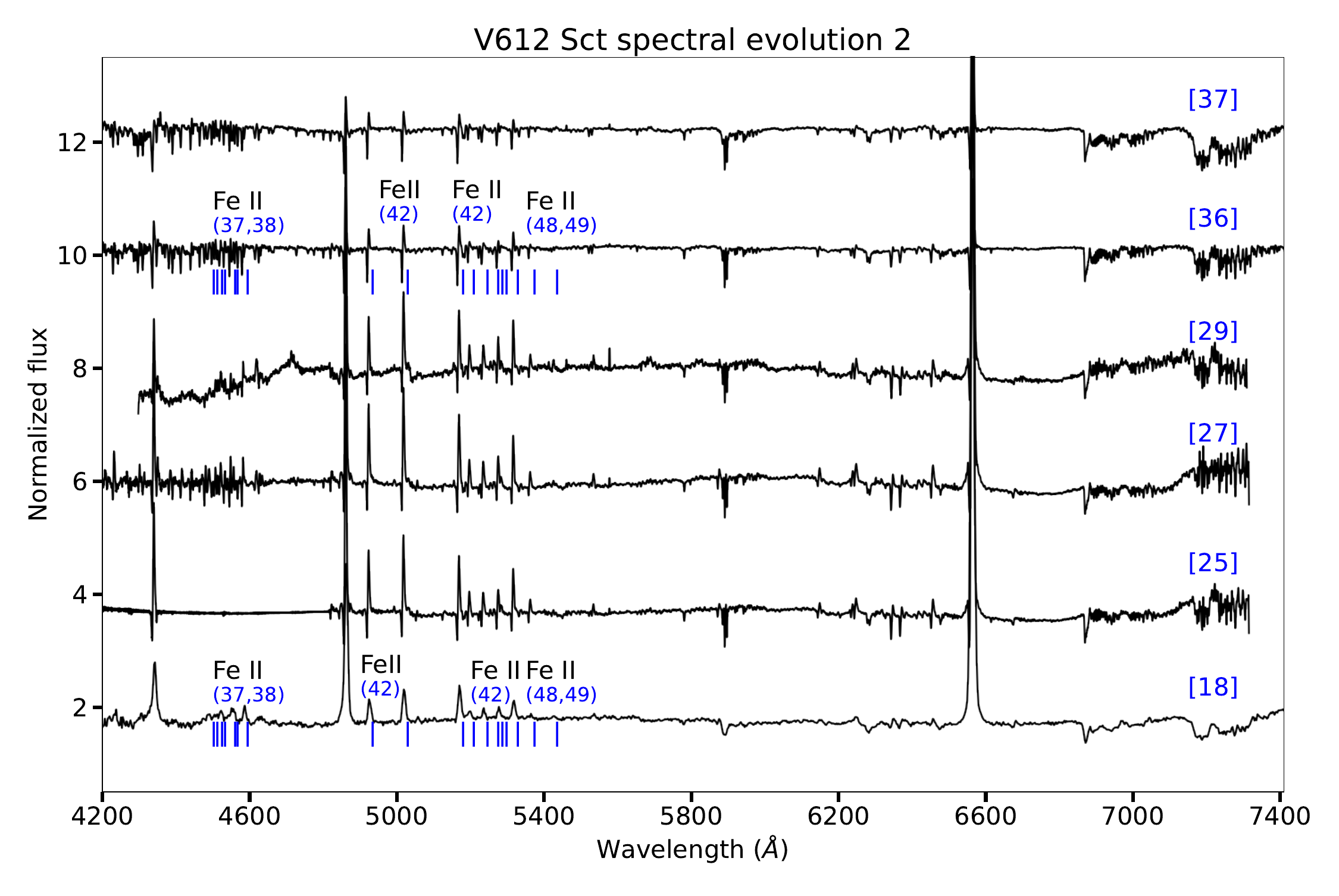}
\caption{The chronological spectral evolution of V612~Sct. The numbers in brackets are days since $t_0$. Colored line identifications are included to help distinguish the spectral features.} 
\label{Fig:spec_2}
\end{center}
\end{figure*}

\begin{figure*}
\begin{center}
  \includegraphics[width=0.8\textwidth]{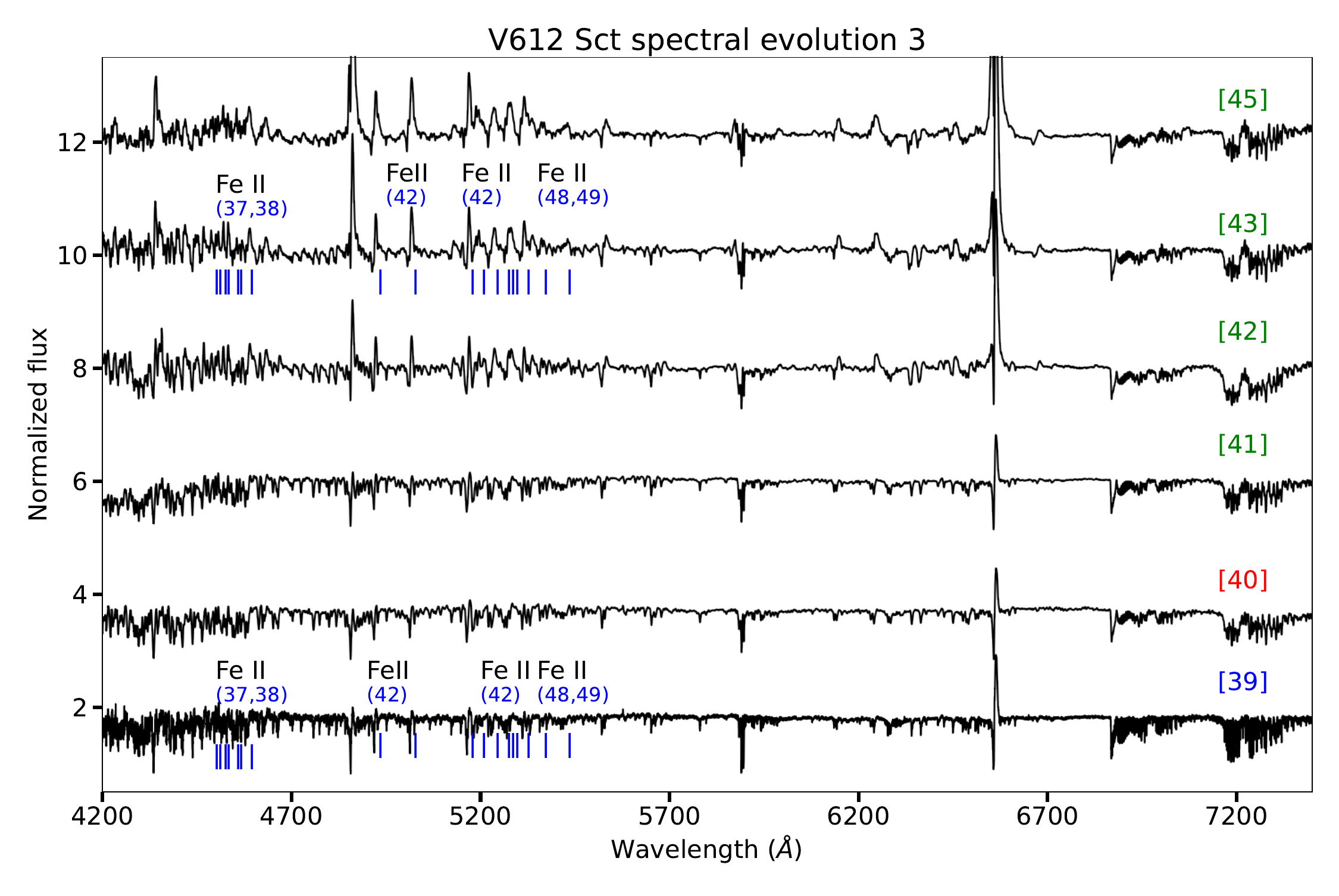}
\caption{The chronological spectral evolution of V612~Sct. The numbers in brackets indicate the days since $t_0$. Colored line identifications are included to help distinguish the spectral features.} 
\label{Fig:spec_3}
\end{center}
\end{figure*}

\begin{figure*}
\begin{center}
  \includegraphics[width=0.8\textwidth]{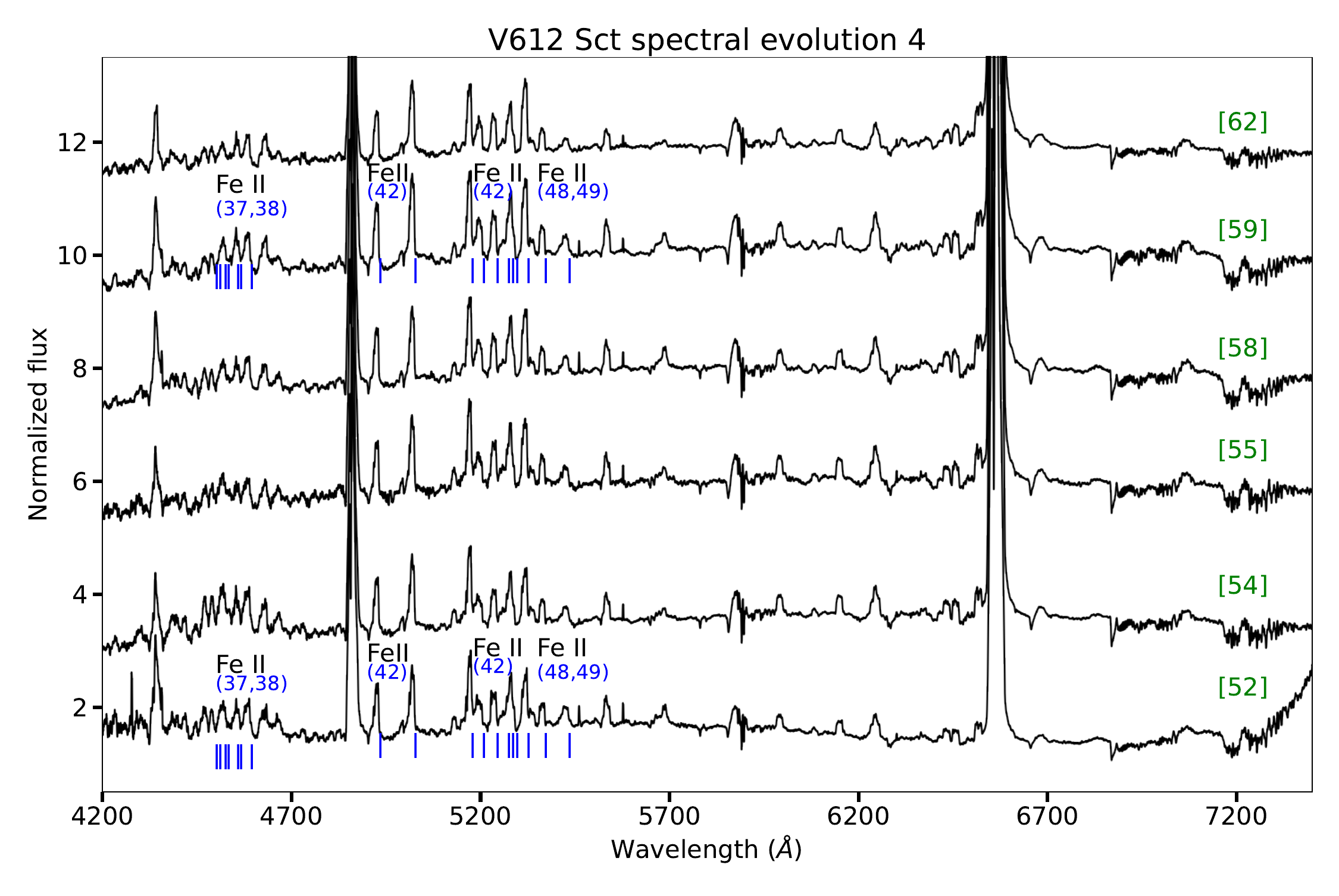}
\caption{The chronological spectra evolution of V612~Sct. The numbers in brackets are since $t_0$. Colored line identifications are included to help distinguish the spectral features.} 
\label{Fig:spec_4}
\end{center}
\end{figure*}

\begin{figure*}
\begin{center}
  \includegraphics[width=0.8\textwidth]{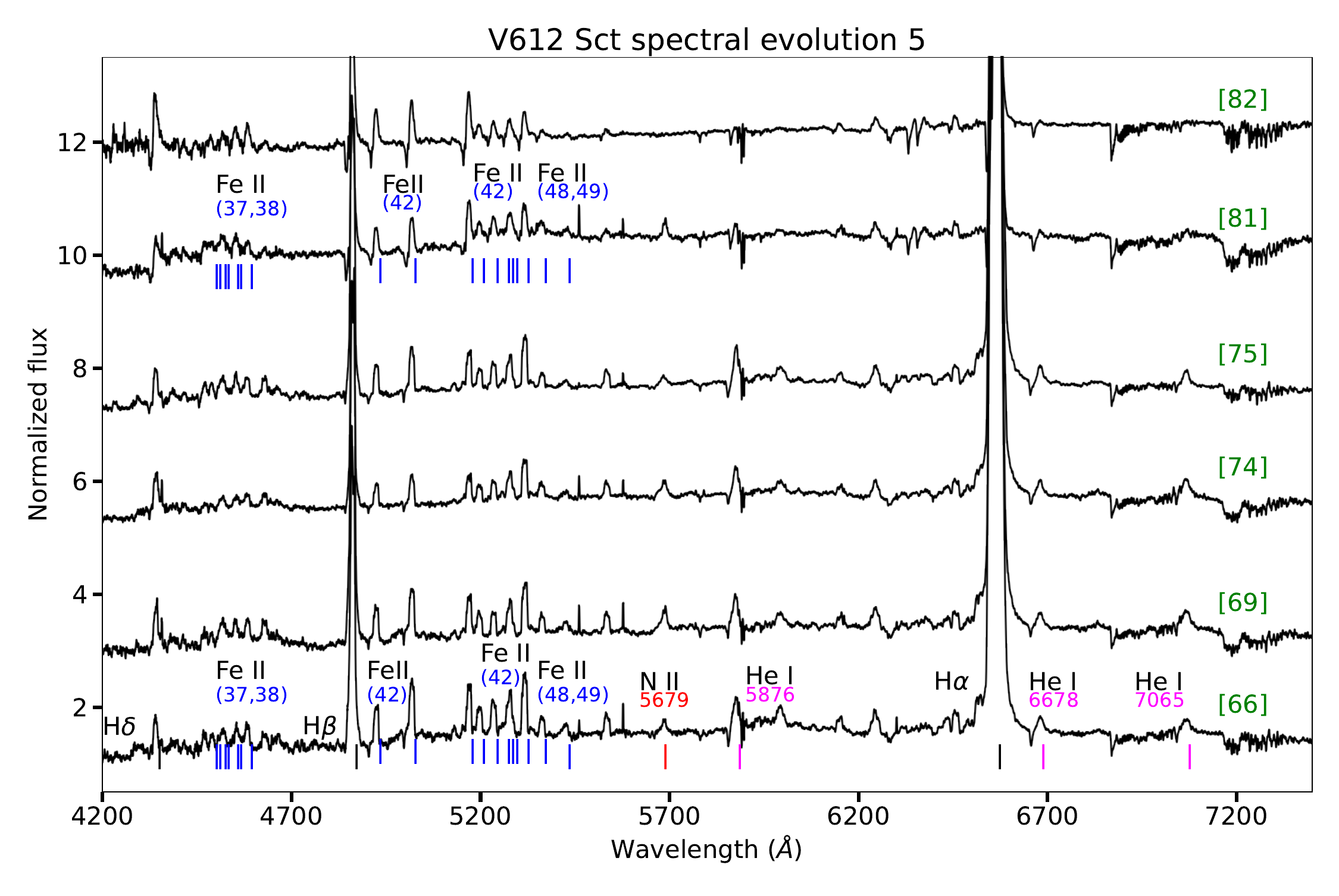}
\caption{The chronological spectra evolution of V612~Sct. The numbers in brackets are days since $t_0$. Colored line identifications are included to help distinguish the spectral features.} 
\label{Fig:spec_5}
\end{center}
\end{figure*}

\begin{figure*}
\begin{center}
  \includegraphics[width=0.8\textwidth]{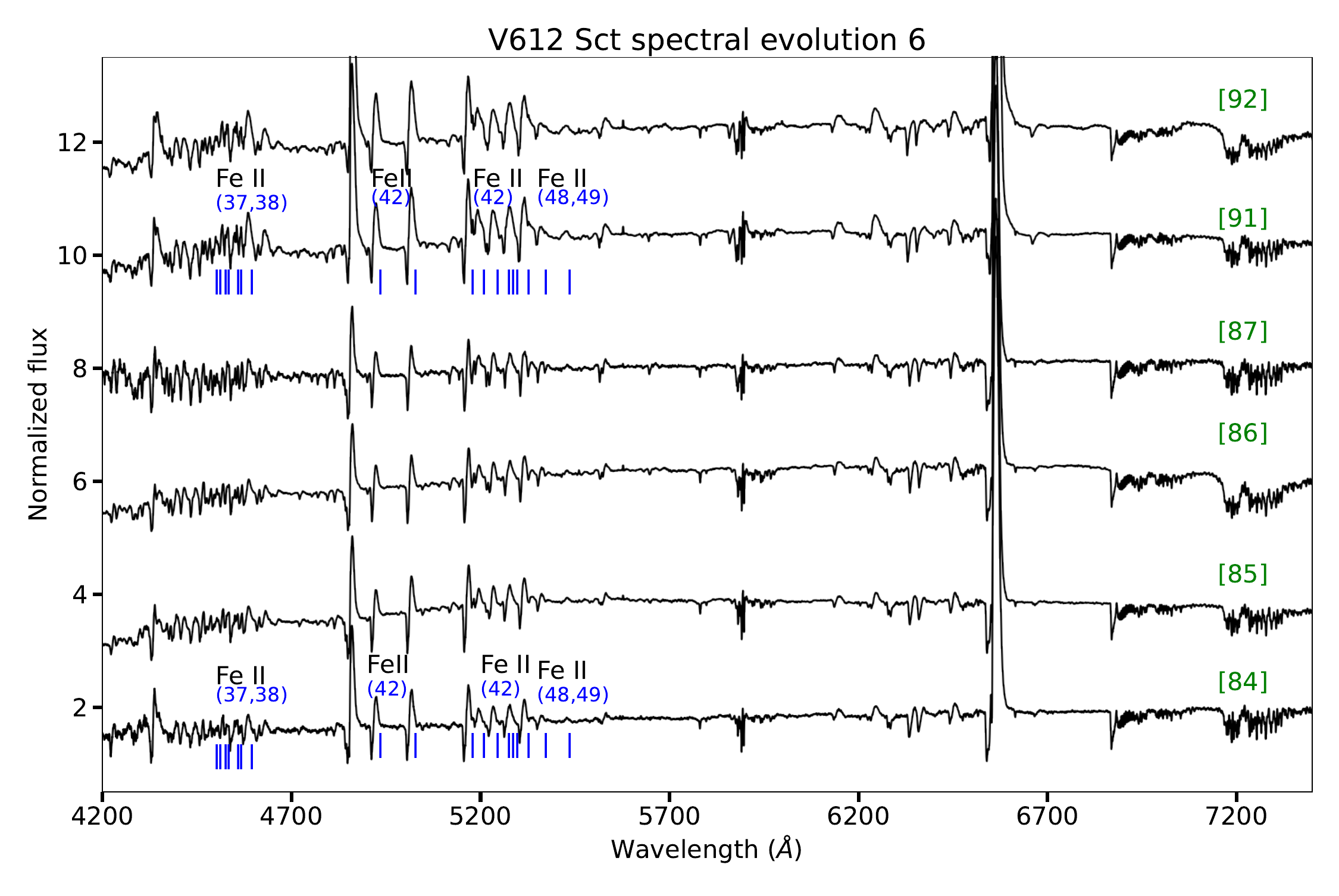}
\caption{The chronological spectra evolution of V612~Sct. The numbers in brackets are days since $t_0$. Colored line identifications are included to help distinguish the spectral features.} 
\label{Fig:spec_6}
\end{center}
\end{figure*}

\begin{figure*}
\begin{center}
  \includegraphics[width=0.8\textwidth]{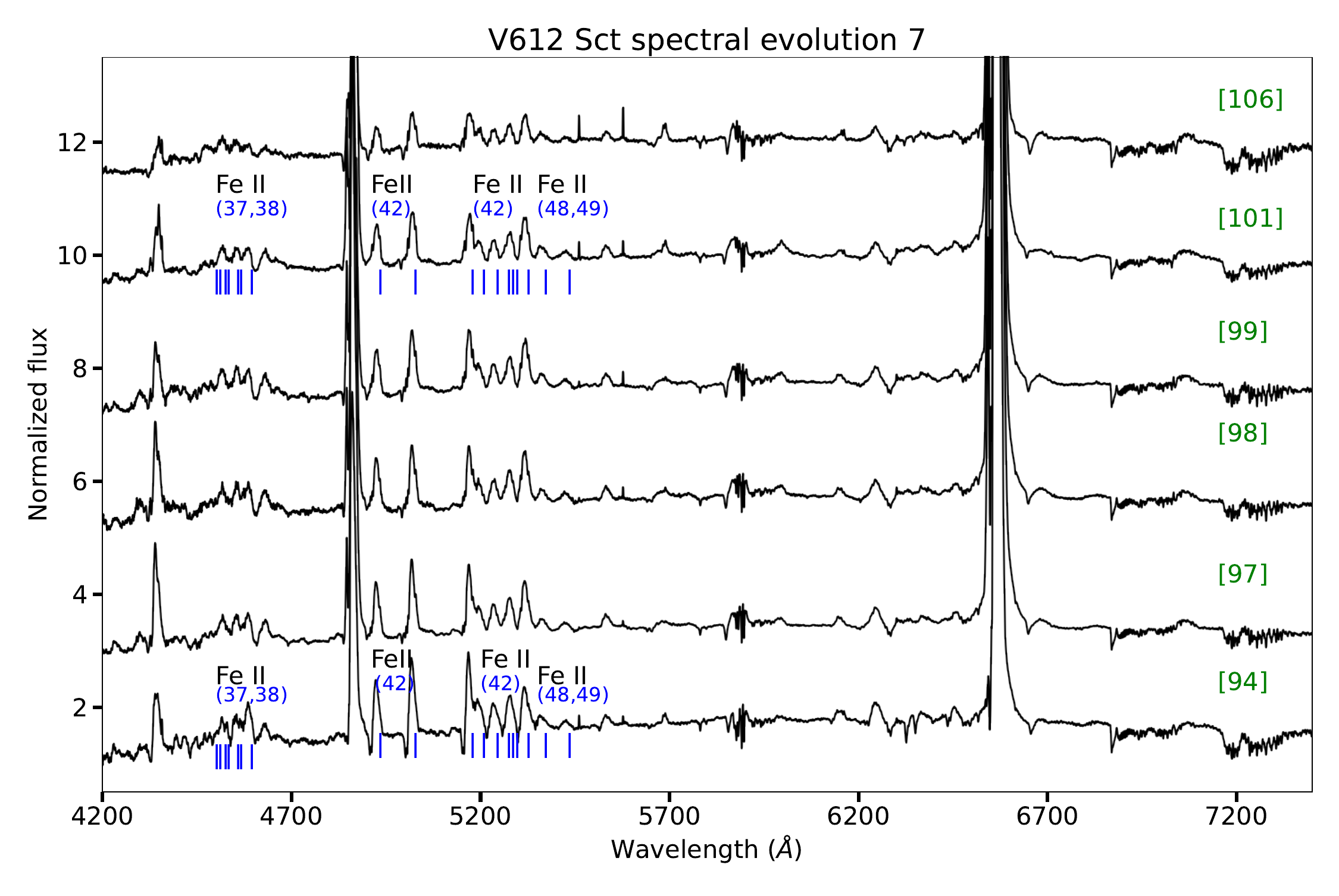}
\caption{The chronological spectra evolution of V612~Sct. The numbers in brackets are days since $t_0$. Colored line identifications are included to help distinguish the spectral features.} 
\label{Fig:spec_7}
\end{center}
\end{figure*}

\begin{figure*}
\begin{center}
  \includegraphics[width=0.8\textwidth]{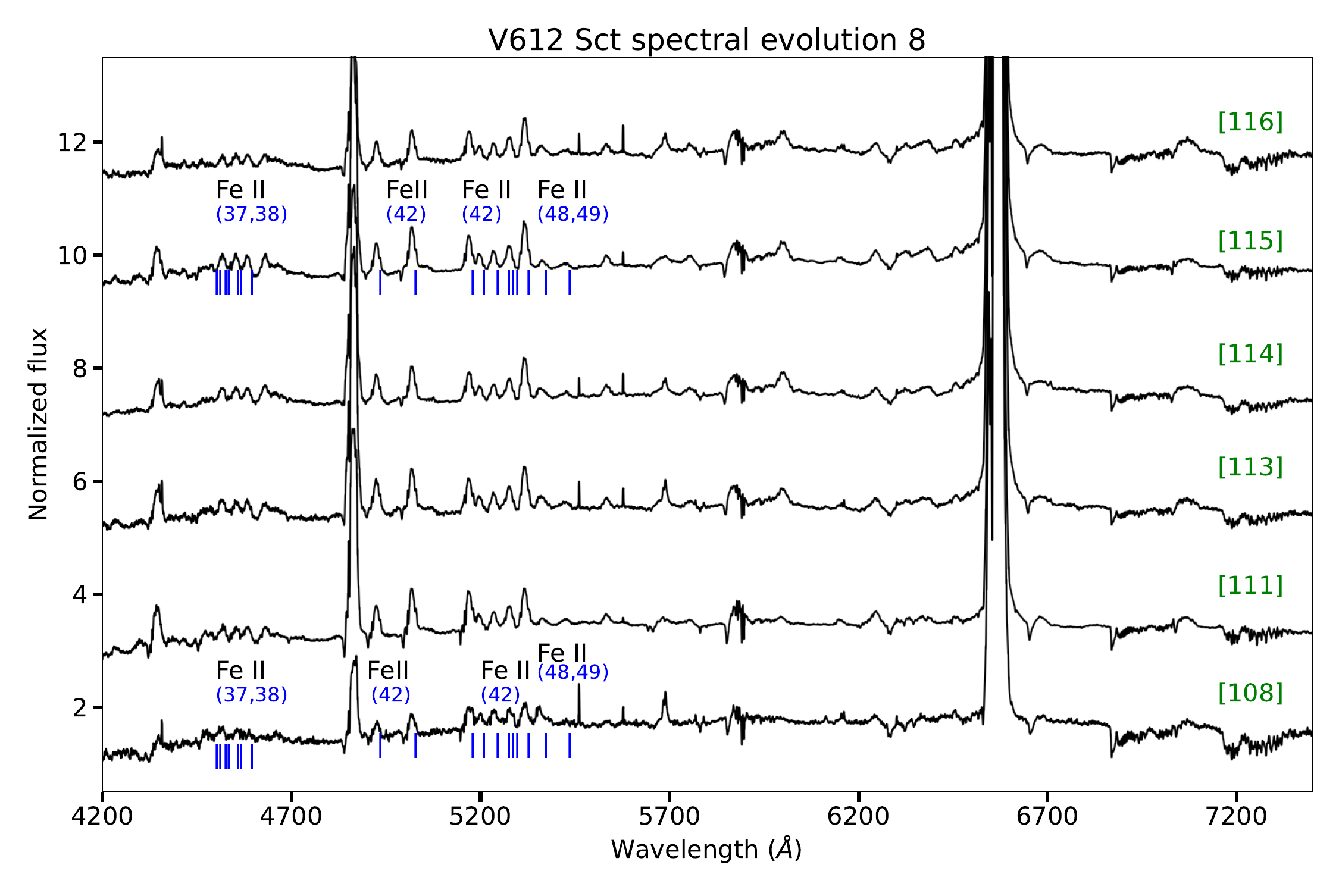}
\caption{The chronological spectra evolution of V612~Sct. The numbers in brackets are days since $t_0$. Colored line identifications are included to help distinguish the spectral features.} 
\label{Fig:spec_8}
\end{center}
\end{figure*}

\begin{figure*}
\begin{center}
  \includegraphics[width=0.8\textwidth]{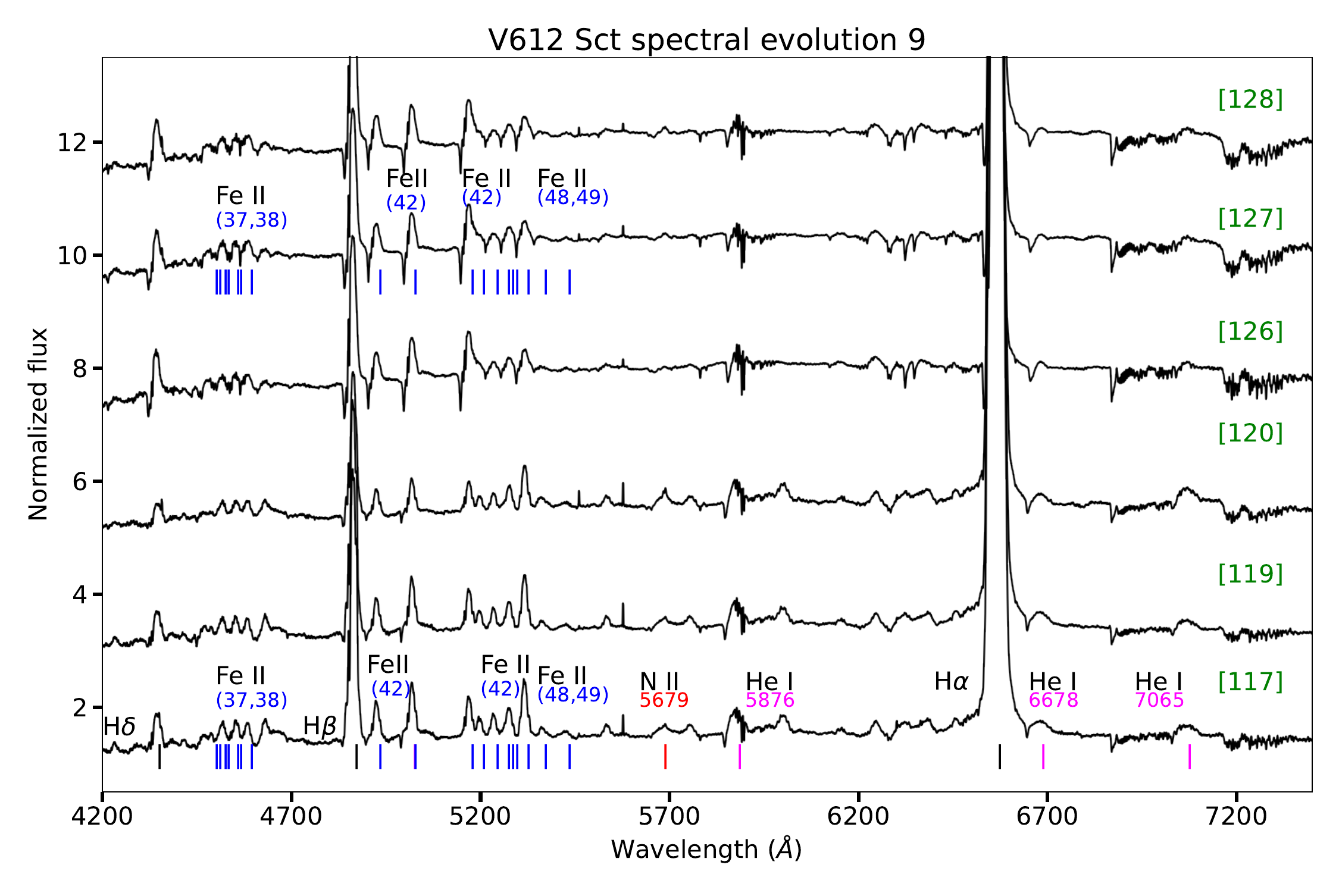}
\caption{The chronological spectra evolution of V612~Sct. The numbers in brackets are days since $t_0$. Colored line identifications are included to help distinguish the spectral features.} 
\label{Fig:spec_9}
\end{center}
\end{figure*}

\begin{figure*}
\begin{center}
  \includegraphics[width=0.8\textwidth]{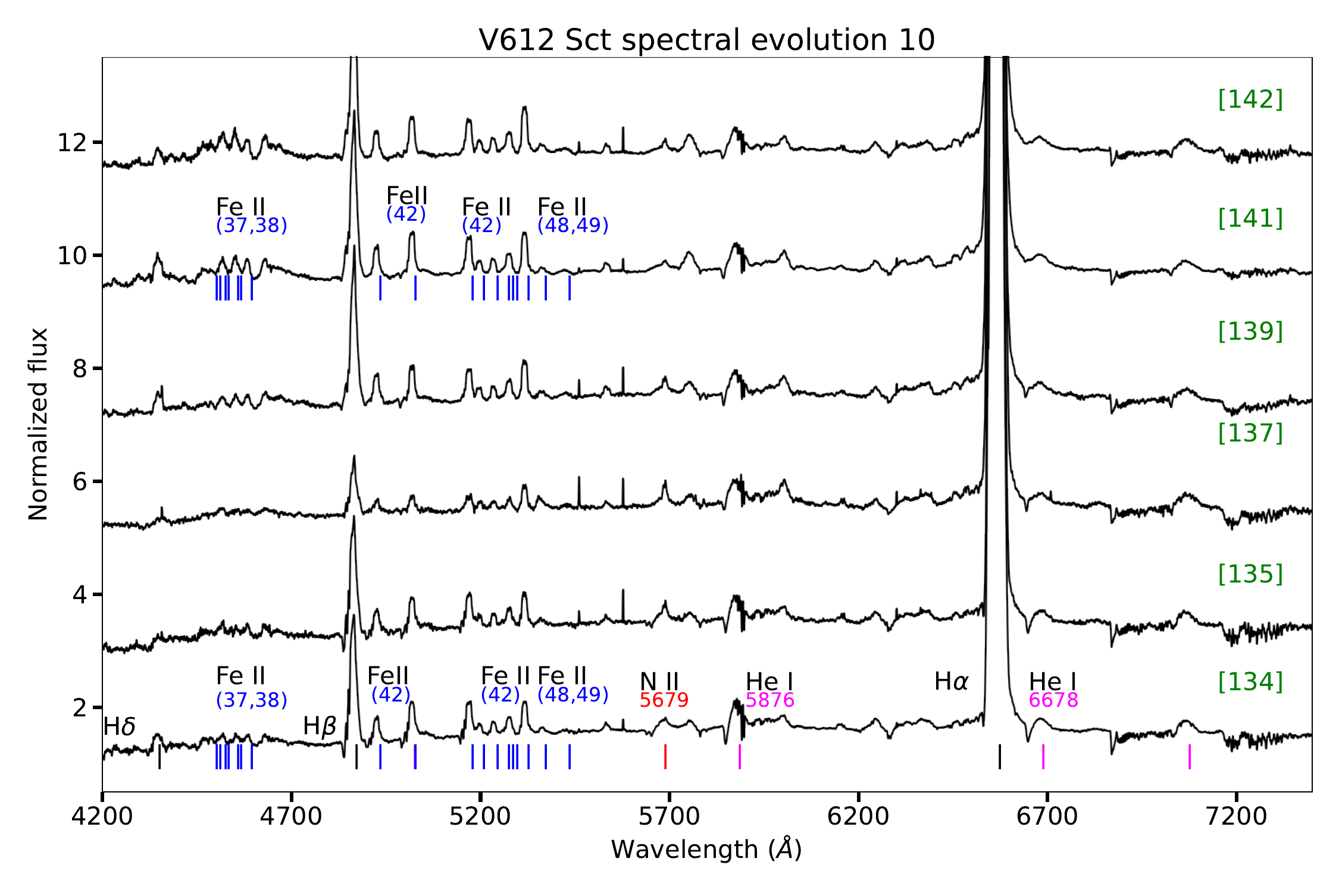}
\caption{The chronological spectra evolution of V612~Sct. The numbers in brackets are days since $t_0$. Colored line identifications are included to help distinguish the spectral features.} 
\label{Fig:spec_10}
\end{center}
\end{figure*}

\begin{figure*}
\begin{center}
  \includegraphics[width=0.8\textwidth]{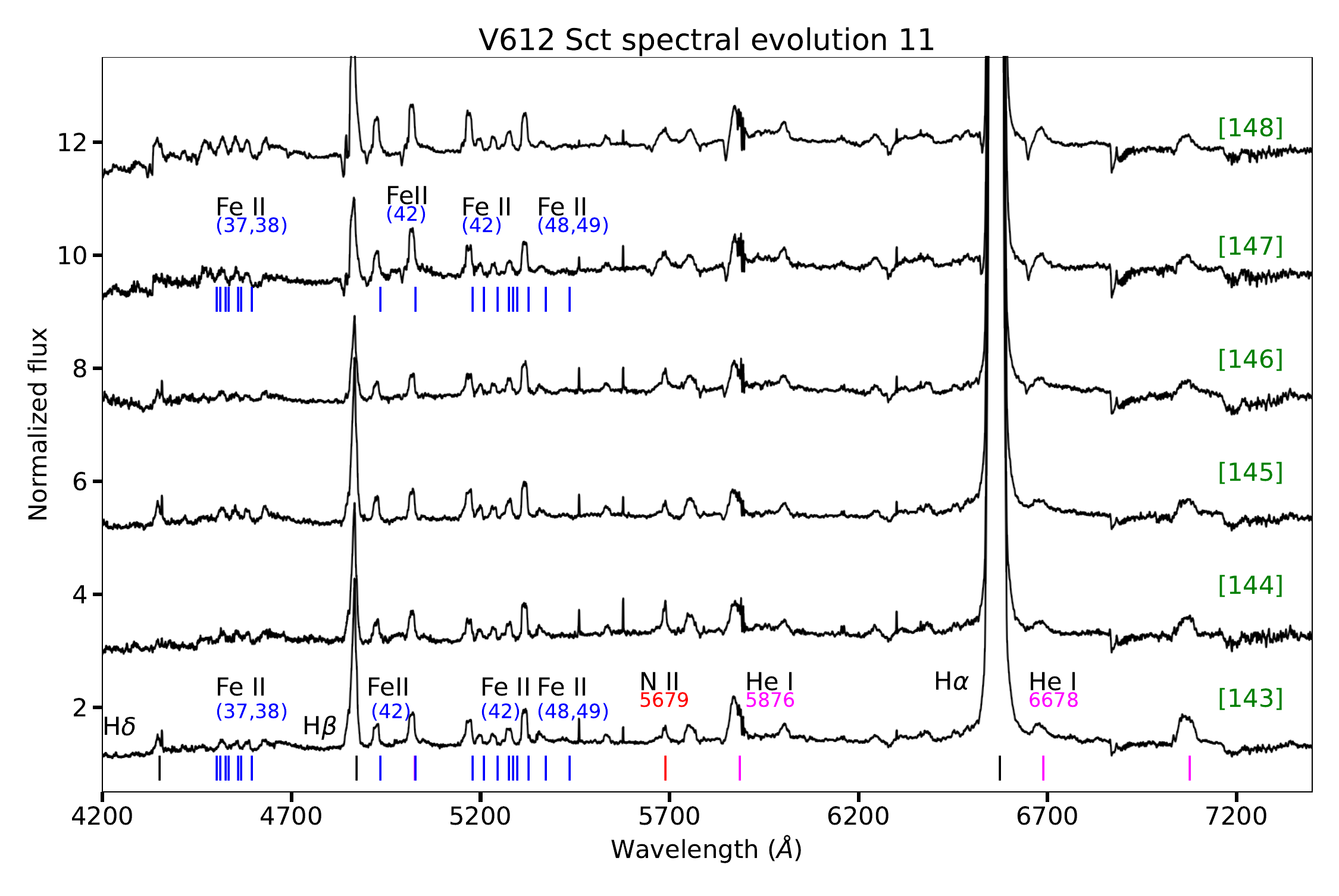}
\caption{The chronological spectra evolution of V612~Sct. The numbers in brackets are days since $t_0$. Colored line identifications are included to help distinguish the spectral features.} 
\label{Fig:spec_11}
\end{center}
\end{figure*}

\begin{figure*}
\begin{center}
  \includegraphics[width=0.8\textwidth]{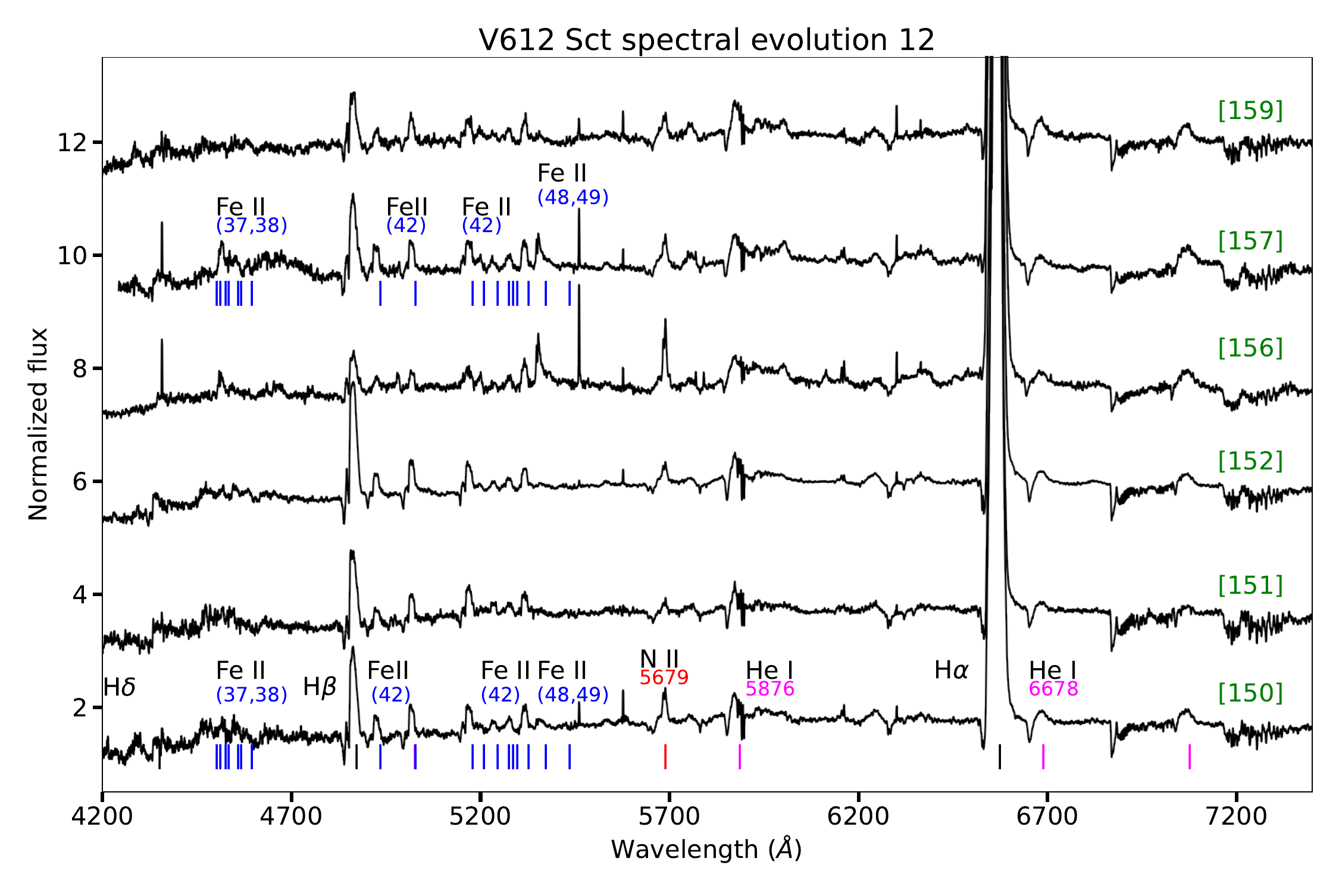}
\caption{The chronological spectra evolution of V612~Sct. The numbers in brackets are days since $t_0$. Colored line identifications are included to help distinguish the spectral features.} 
\label{Fig:spec_12}
\end{center}
\end{figure*}

\begin{figure*}
\begin{center}
  \includegraphics[width=0.8\textwidth]{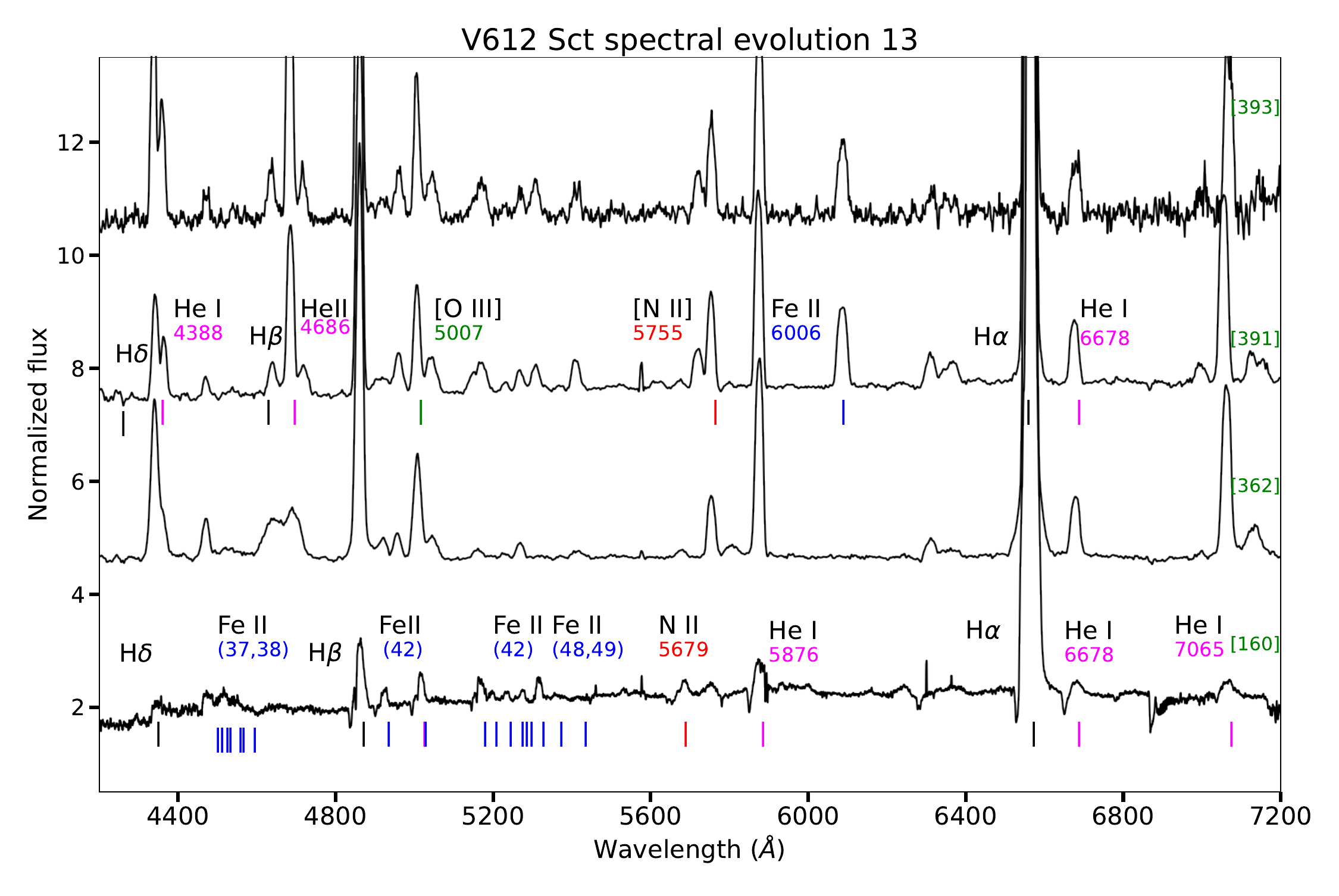}
\caption{The chronological spectra evolution of V612~Sct. The numbers in brackets are days since $t_0$. Colored line identifications are included to help distinguish the spectral features.} 
\label{Fig:spec_13}
\end{center}
\end{figure*}

\begin{figure*}
\begin{center}
  \includegraphics[width=1.0\textwidth]{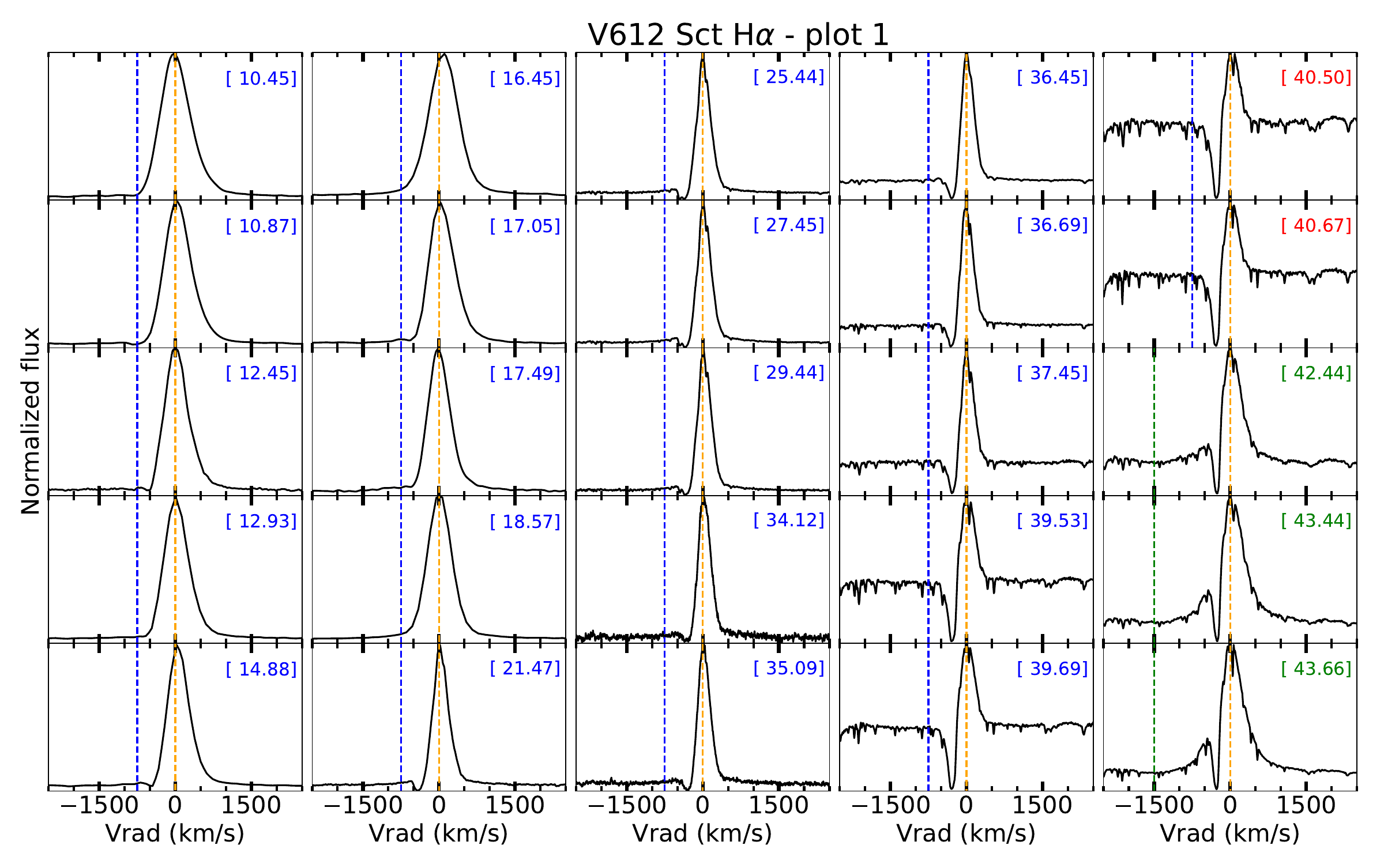}
\caption{The evolution of the H$\alpha$ line profiles. The numbers between brackets are days since $t_0$. The days highlighted in blue are of spectra taken before peak brightness, while days highlighted in green are that of spectra taken after peak brightness. The red, blue, and green dashed lines represent $v_{\mathrm{rad}}$ = 0\,km\,s$^{-1}$, $v_{\mathrm{rad}}$ = $750$\,km\,s$^{-1}$, $v_{\mathrm{rad}}$ = $-1500$\,km\,s$^{-1}$, respectively.} 
\label{Fig:line_profile_1}
\end{center}
\end{figure*}

\begin{figure*}
\begin{center}
  \includegraphics[width=1.0\textwidth]{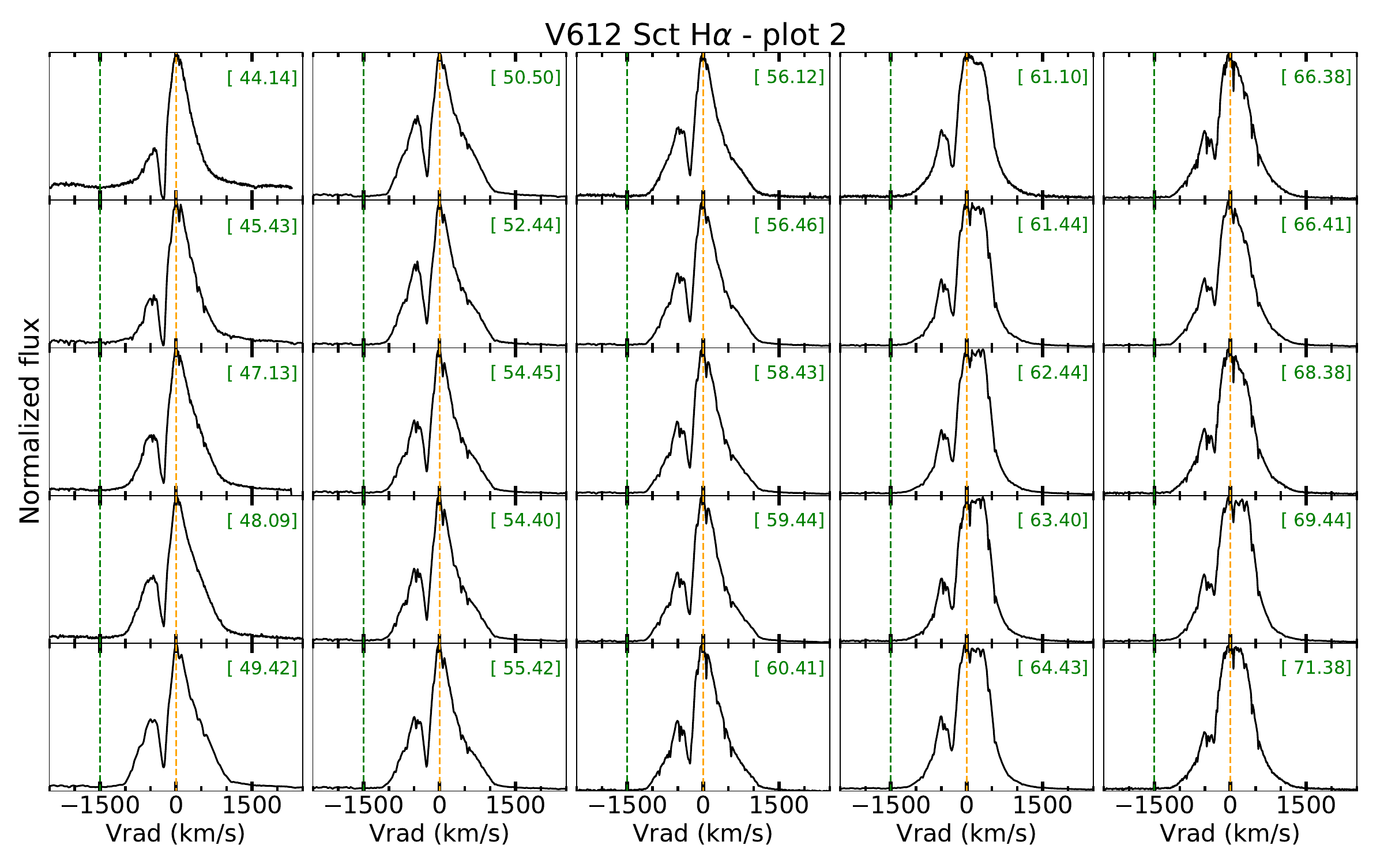}
\caption{The evolution of the H$\alpha$ line profiles. The numbers between brackets are days since $t_0$. The days highlighted in blue are of spectra taken before peak brightness, while days highlighted in green are that of spectra taken after peak brightness. The red, blue, and green dashed lines represent $v_{\mathrm{rad}}$ = 0\,km\,s$^{-1}$, $v_{\mathrm{rad}}$ = $750$\,km\,s$^{-1}$, $v_{\mathrm{rad}}$ = $-1500$\,km\,s$^{-1}$, respectively.} 
\label{Fig:line_profile_2}
\end{center}
\end{figure*}

\begin{figure*}
\begin{center}
  \includegraphics[width=1.0\textwidth]{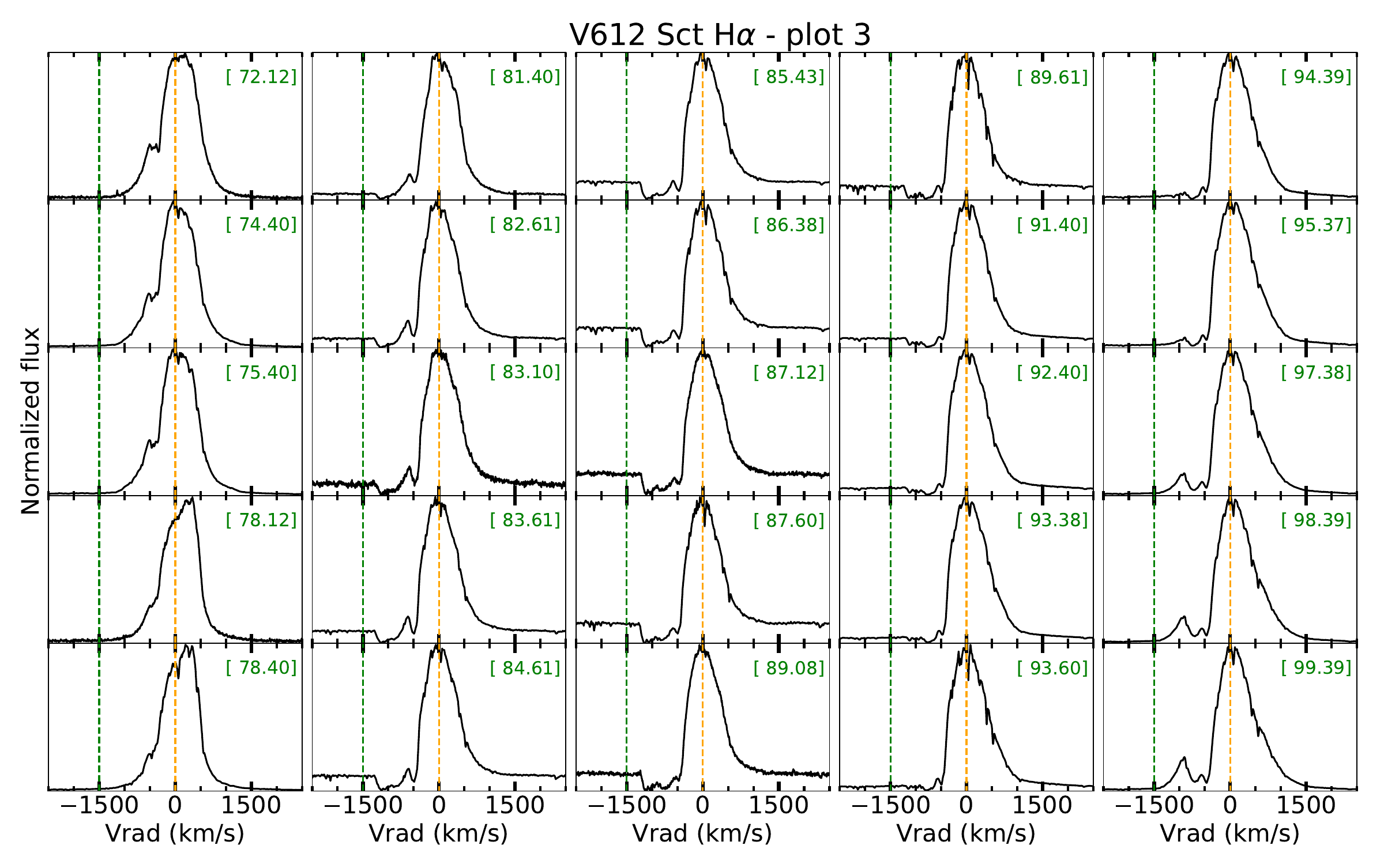}
\caption{The evolution ofthe H$\alpha$ line profiles. The numbers between brackets are days since $t_0$. The days highlighted in blue are of spectra taken before peak brightness, while days highlighted in green are that of spectra taken after peak brightness. The red, blue, and green dashed lines represent $v_{\mathrm{rad}}$ = 0\,km\,s$^{-1}$, $v_{\mathrm{rad}}$ = $750$\,km\,s$^{-1}$, $v_{\mathrm{rad}}$ = $-1500$\,km\,s$^{-1}$, respectively.} 
\label{Fig:line_profile_3}
\end{center}
\end{figure*}

\begin{figure*}
\begin{center}
  \includegraphics[width=1.0\textwidth]{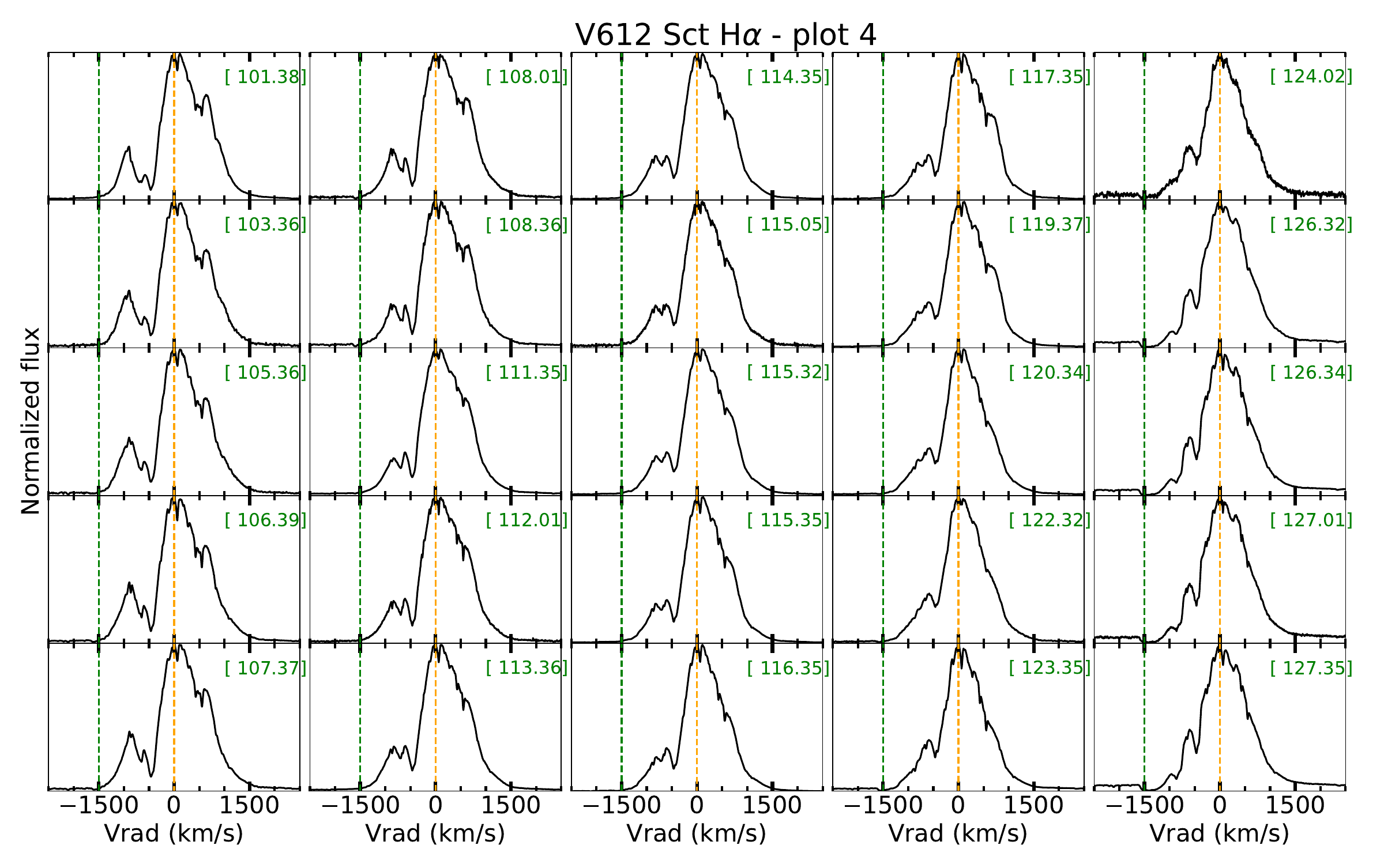}
\caption{The evolution of the H$\alpha$ line profiles. The numbers between brackets are days since $t_0$. The days highlighted in blue are of spectra taken before peak brightness, while days highlighted in green are that of spectra taken after peak brightness. The red, blue, and green dashed lines represent $v_{\mathrm{rad}}$ = 0\,km\,s$^{-1}$, $v_{\mathrm{rad}}$ = $750$\,km\,s$^{-1}$, $v_{\mathrm{rad}}$ = $-1500$\,km\,s$^{-1}$, respectively. } 
\label{Fig:line_profile_4}
\end{center}
\end{figure*}

\begin{figure*}
\begin{center}
  \includegraphics[width=1.0\textwidth]{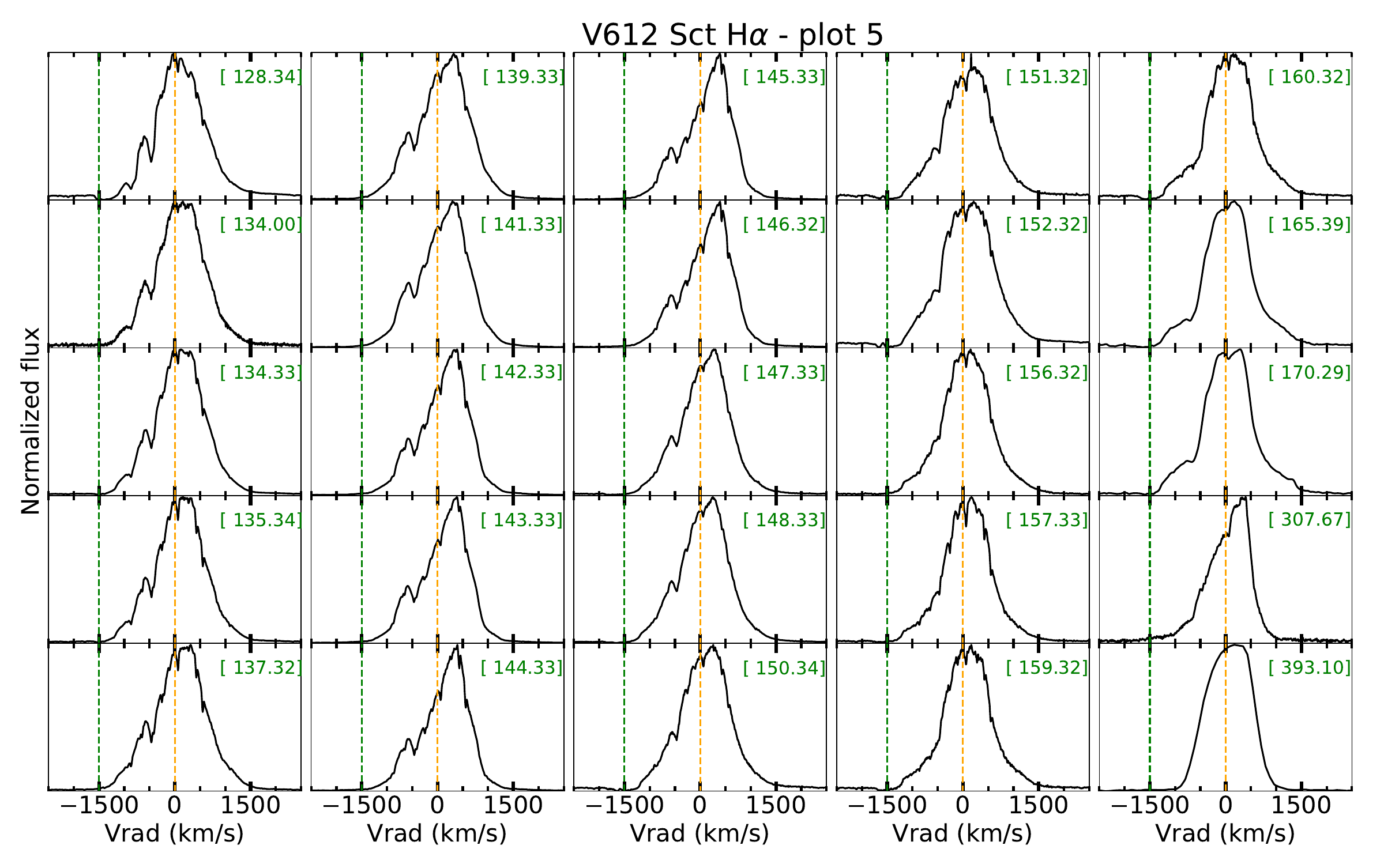}
\caption{The evolution of the H$\alpha$ line profiles. The numbers between brackets are days since $t_0$. The days highlighted in blue are of spectra taken before peak brightness, while days highlighted in green are that of spectra taken after peak brightness. The red, blue, and green dashed lines represent $v_{\mathrm{rad}}$ = 0\,km\,s$^{-1}$, $v_{\mathrm{rad}}$ = $750$\,km\,s$^{-1}$, $v_{\mathrm{rad}}$ = $-1500$\,km\,s$^{-1}$, respectively.} 
\label{Fig:line_profile_5}
\end{center}
\end{figure*}

\end{document}